\documentclass[floatfix,aps,preprint,nofootinbib]{revtex4-2}

\usepackage[paperwidth=210mm,paperheight=297mm,centering,hmargin=2cm,vmargin=2.5cm]{geometry}

\usepackage[utf8]{inputenc}
\usepackage{amsfonts}
\usepackage{color}

\usepackage{caption}
\usepackage{booktabs}

\usepackage{amssymb}
\usepackage{amsmath}
\usepackage{stmaryrd}
\usepackage{latexsym}

\usepackage{xcolor}
\definecolor{myurlcolor}{HTML}{08457E}
\definecolor{mylinkcolor}{HTML}{2A52BE}
\definecolor{mycitecolor}{HTML}{E30022}

\usepackage{graphicx}
\usepackage{cancel}
\usepackage{braket}
\usepackage{siunitx}
\usepackage{xspace}
\usepackage{tcolorbox}
\usepackage{subcaption}
\usepackage{mathtools}
\usepackage{tabularx}

\usepackage{floatrow}
\usepackage{float}
\usepackage[colorlinks, linkcolor=mylinkcolor, citecolor=mycitecolor, urlcolor=myurlcolor, linktocpage=true]{hyperref}

\def\equationautorefname~#1\null{(#1)\null}
\def\tableautorefname~#1\null{(#1)\null}
\def\figureautorefname~#1\null{(#1)\null}
\def\sectionautorefname~#1\null{(#1)\null}

\let\origref\autoref
\def\autoref#1{\textbf{\origref{#1}}}

\let\origcite\cite
\def\cite#1{\textbf{\origcite{#1}}}

\usepackage{multirow}
\usepackage{tikz}
\usepackage{pgfplots}
\pgfplotsset{compat=1.15}

\usepackage{titlesec}
\titleformat*{\section}{\centering\small\bfseries\scshape}
\titleformat*{\subsection}{\small\bfseries\scshape}
\titleformat*{\subsubsection}{\small\bfseries\scshape}

\let\oldsqrt\sqrt
\def\sqrt{\mathpalette\DHLhksqrt}
\def\DHLhksqrt#1#2{%
	\setbox0=\hbox{$#1\oldsqrt{#2\,}$}\dimen0=\ht0
	\advance\dimen0-0.4\ht0
	\setbox2=\hbox{\vrule height\ht0 depth -\dimen0}%
	{\box0\lower0.4pt\box2}}

\def\jcap{Journal of Cosmology and Astroparticle Physics}
\def\prd{Physical Review D}

\def\physrep{Phys. Rep.}

\newfloatcommand{capbtabbox}{table}[][\FBwidth]

\begin{document}

\title{\boldmath Stability Analysis of the Cosmological Dynamics of $O(D,D)$-complete Stringy Gravity\vspace{1cm}}

\author{A. Sava{\c s} Arapo{\u g}lu}
\email{arapoglu@itu.edu.tr}
\author{Sermet {\c C}a{\u g}an}
\email{cagans@itu.edu.tr}
\affiliation{Istanbul Technical University, Department of Physics, 34469 Maslak, Istanbul, Turkey}

\author{Aybike \c{C}atal-\"{O}zer}
\email{ozerayb@itu.edu.tr}
\affiliation{Istanbul Technical University, Department of Mathematics, 34469 Maslak, Istanbul, Turkey \vspace{2cm}}

\begin{abstract}
The massless fields in the universal NS-NS sector of string theory form  $O(D, D)$ multiplets of Double Field Theory, which is a theory that provides a T-duality covariant formulation of supergravity, leading to a stringy modification of General Relativity. 
In this framework, it is possible to write down the extensions of the Einstein field equations and the Friedmann equations in such a way that the coupling of gravitational and matter sectors is dictated by the $O(D, D)$ symmetry universally. In this paper, we obtain the autonomous form of the $O(D, D)$-complete Friedmann equations, find the critical points and perform their stability analysis. We also include the phase portraits of the system. Cosmologically interesting cases of scalar field, radiation, and matter are separately considered and compared with the Chameleon models in a similar setting. Accelerating phases and the conditions for their existence are also given for such cases.
\end{abstract}

\maketitle
\clearpage
\raggedbottom

\section{Introduction}
\label{sec:intro}
Alternative theories of gravity to general relativity have become an active field of research, especially in the last 30 years, for both observational and conceptual reasons. When observations at cosmological scales are tried to be understood within the framework of general relativity, the fact that unknown matter-energy components such as dark matter and dark energy have to be included in the models of the universe is one of the most important reasons for this field to become active. Moreover, some tensions/incompatibilities of the $\Lambda$CDM model built within the framework of general relativity, which has recently emerged as a result of observations made at various scales, has made it a kind of necessity to address the theoretical framework of the cosmological standard model. Considering all these about the early universe processes suggests that alternative theories of gravity should be examined together with consistent quantum gravity theory candidates such as string theory. The low energy limits of string theories provide an important alternative set of gravitational theories for understanding how the early universe processes and the present large-scale structure of the universe emerged with such processes. Cosmology, thus, offers a very promising window for examining the observational consequences of string theory and is also easier in the sense that studying only time-dependent backgrounds is simpler than considering space and time coordinates together.

`Stringy gravity' emerging from the low energy limit of string theory has more fundamental gravitational fields than general relativity: in addition to the metric $g_{\mu\nu}$, there are also other massless modes of the NS-NS sector, which are the dilaton and the two-form field $B_{\mu\nu}$. The dilaton and B-field are, in general, treated as moduli to be dynamically stabilized by some means in the quest for superstring vacua \citep{damour-poly94-1,damour-poly94-2}, and the dilaton is included in the framework of SUGRA cosmology \citep{gasper2003, vafa1992, gasper2008, veneziano91, gasper92}.  A crucial difficulty with the inclusion of the dilaton and/or B-field in the search for cosmological solutions is how to couple these stringy gravitational fields with matter. The usual strategy is to minimally couple the metric to the matter sector; but the couplings of dilaton and B-field to matter are omitted since these couplings lead to some difficulties concerning observational restrictions, like solar system constraints \citep{damour-poly94-2}. Actually, coupling dilaton (or any light scalar) to the matter sector is highly restricted by similar constraints coming from different scales. A way of evading these problems caused by scalar fields is to introduce some `screening' mechanism to build cosmological models consistent with observations \citep{brax2013}. The coupling of the scalar field to matter, together with the form of the scalar field potential may lead to a rich phenomenology, including non-perturbative phenomena such as spontaneous scalarization also \citep{damour1993, fethi2024} and references therein.

From this point of view, it would be crucial to determine how the scalar field couples to matter, not just on a phenomenological ground, but in a way dictated by basic symmetry principles of the underlying (quantum) theory of gravity. Observational implications resulting from this coupling, imposed by fundamental symmetry principles of the theory, would also be critical to test the consistency of the theory. 
In string theory,  the dilaton's interaction with matter is usually interpreted as an ad hoc addition, necessitated by the conformal anomaly and specific field content. Double Field Theory (DFT) changes this perspective by introducing an   $O(D, D)$ T-duality covariant action that reduces to the low energy action for the universal sector of string theory, when a constraint, again  $O(D, D)$ covariant, is imposed \citep{hull-zwiebach2009}.   This is achieved by introducing $D$ coordinates conjugate to string winding modes in addition to the $D$ space-time coordinates conjugate to momentum modes, in such a way that they form an  $O(D, D)$ mutliplet.  The dynamics of the theory is governed by  field equations, which involves a generalized Ricci tensor and a generalized Ricci scalar. The generalized Ricci tensor $\mathcal{R}(\mathcal{H}, d)$ involves the generalized dilaton $d$ and the generalized metric $\mathcal{H}$, which comprises of the Riemannian metric and the antisymmetric B-field \citep{hohmetal2010-1, hohmetal2010-2}. 
DFT has also been extended so as to incorporate matter fields and in this framework it is possible to write the generalization of Einstein Field Equations by generalizing not just the Einstein tensor but also the energy-momentum tensor. This makes it possible to fix the coupling of matter to metric, the B-field and, most importantly for the purposes of this paper, the dilaton field  by requiring the interactions to be consistent with the  $O(D, D)$ T-duality symmetry. The generalized equations describing the interaction between the gravitational and the matter sectors are called the Einstein Double Field Equations (EDFEs) \citep{angus2018}, and they may be considered as $O(D, D)$-completion of equations of GR (For the earlier discussions on the cosmological implications of the  $O(D, D)$ symmetry include \citep{wu2014, ma2014, branden2018-1, branden2018-2, brandenetal2019, branden-bernardo2019}). Solar system tests and the post-Newtonian DFT analysis is considered in \citep{Choi:2022srv}.
The generalized Friedmann equations following from this comprehensive $O(D, D)$-symmetric framework, called the $O(D, D)$-complete Friedmann Equations (OFEs), is derived in \citep{Angus2020} and also considered in \citep{Lee:2023} (For the $\alpha^\prime$-corrected $O(D, D)$ cosmology considerations see \citep{Bernardo:2019bkz, Bernardo:2022nex, Quintin:2021eup, Conzinu:2023fth}). This $O(D, D)$-symmetric formulation modifies SUGRA cosmology in such a way that whenever the dilaton is stabilized, the standard Friedmann equations are obtained not only for a radiation-dominated universe as in the traditional string cosmology but also in the presence of any matter sources. 

Our primary motivation in this work is to examine the viability of the coupling of dilaton to the matter sector dictated by the $O(D, D)$ symmetry of the Double Field Theory (DFT) in the cosmological setting. In this study, we compare the $O(D, D)$-extended system with the so-called ``Chameleon Models'' where the form of the potential of the scalar field and the coupling of the dilaton to matter sector may lead to a special type of a screening effect for the scalar field. In the analysis of $O(D, D)$ system and comparison with Chameleon model, we use the dynamical system approach \citep{bohmer2018}. This allows to set up a framework that takes into account plausible cosmic situations and explores the viability of the models \citep{bernardo2020}.

The organization of the paper is as follows. In section 2, $O(D, D)$-complete Friedmann Equations are put into the form of a system of autonomous equations. In section 3, the critical points are determined and their stability analysis is considered, together with phase portraits. In section 4, some cosmologically interesting cases for phenomenological applications are considered and possible implications are mentioned. In section 5, $O(D, D)$-complete cosmological models are compared with another string-inspired phenomenological model, called Chameleon gravity \citep{chameleon, chameleon2, chameleon3}. In section 6, stability analysis of the case $k\ne0$ is considered. We conclude with some comments and possible future work in section 7.

\section{Autonomous Equation System for $O(D, D)$-complete Cosmology}

The $O(D, D)$ complete Friedmann equations was constructed in Ref.~\citep{Angus2020} by imposing the most general ansatz for a homogeneous and isotropic cosmological background in four dimensions. The ansatz is dictated by isometries of DFT and substituting them in  Einstein Double Field Equations yields $O(D, D)$-complete Friedmann Equations as given in \citep{Angus2020}. Here, we aim to perform a linear stability analysis on the $O(D, D)$-complete Friedmann equations of a spatially flat ($k=0$) FRW Universe, so we rearrange these equations using the cosmic gauge by setting the lapse function, $N(t)$, to unity.  As a result, we arrive at the following set of $O\left(D,D\right)$-complete Friedmann equations,
\begin{subequations}
\begin{align}
    \label{eq:odd_feq1}\frac{8\pi G}{3}\rho e^{2\phi} + \frac{h^{2}}{12 a^{6}} &= H^{2} - 2\dot{\phi}H + \frac{2}{3}\dot{\phi}^{2}\\
    \label{eq:odd_feq2}\frac{4\pi G}{3}\left(\rho + 3p\right)e^{2\phi} + \frac{h^{2}}{6a^{6}} &= -H^{2} - \dot{H} + \dot{\phi}H - \frac{2}{3}\dot{\phi}^{2} + \ddot{\phi}\\
    \label{eq:odd_feq3}\frac{8\pi G}{3}\left(\rho e^{2\phi} - \frac{1}{2}T_{(0)}\right) &= -H^{2} - \dot{H} + \frac{2}{3}\ddot{\phi}.
\end{align}
\end{subequations}
To have a complete set of equations, we can write the continuity equation as below;
\begin{align}
    \label{eq:continutiy}\dot{\rho} + 3H\left(\rho + p\right) + \dot{\phi}T_{(0)}e^{-2\phi} = 0
\end{align}
 where $\rho$ and $p$ are defined in analogy with General Relativity. The analytic solution of the continuity equation~\eqref{eq:continutiy} can be found by considering non-interacting perfect fluids with constant equation of state parameters, $\omega$, and the parameter $\lambda$ describing the coupling of dilaton to matter:
\begin{align}
    \rho = \rho_{0}a^{-3\left(1+\omega\right)}e^{-\lambda\phi}
\end{align}
where,
\begin{align}
    \label{eq:equations_of_state}
    \omega \equiv p/\rho,\qquad\lambda \equiv T_{(0)}\left(\rho e^{2\phi}\right)^{-1}
\end{align}
In \citep{Choi:2022srv}, an alternative identification for energy density and pressure is used, where $\bar{\rho} \equiv-e^{-2 \phi} K_t^t$ is the mass density and its spatial counterpart is $\bar{\mathbf{p}} \equiv \frac{1}{3} e^{-2 \phi}\left(K_\mu^\mu-K_t^t\right)$. These alternative identifications lead to changes in the parameters that we have used here as follows:
\begin{align}
\bar{\rho} = \rho - \frac{1}{2}T_{\left(0\right)}e^{-2\phi},\quad \bar{\mathbf{p}} = p + \frac{1}{2}T_{\left(0\right)}e^{-2\phi}
\end{align}
which leads to the following relations between the equations of state parameters,
\begin{align}
\omega = \dfrac{\bar{\omega} - \dfrac{1}{2}\bar{\lambda}}{1 + \dfrac{1}{2}\bar{\lambda}}, \quad \lambda = \dfrac{\bar{\lambda}}{1 + \dfrac{1}{2}\bar{\lambda}}.
\end{align}
However, we continue the analysis sticking to the parametrization in eq.~\eqref{eq:equations_of_state}.

An autonomous equation system can be constructed starting with defining the boundaries of the phase space by dividing both sides of the eq.~\eqref{eq:odd_feq1} by $H^{2}$ as,
\begin{align}
    \label{eq:constraint}\frac{8\pi G}{3H^{2}}\rho e^{2\phi} + \frac{h^{2}}{12 a^{6}H^{2}} + 2\frac{\dot{\phi}}{H} - \frac{2}{3}\left(\frac{\dot{\phi}}{H}\right)^{2} = 1
\end{align}
for which the eq.~\eqref{eq:constraint} now gives the constraint of our phase space.
We use the constraint equation as our reference to define dimensionless phase space dynamical variables, that ultimately will describe the behavior of the kinetic term of the scalar field, the flux term, and the energy density term in our equations.
\begin{align}
    \label{eq:variables1}x \equiv \frac{\dot{\phi}}{H}\quad\quad y \equiv \frac{h}{2\sqrt{3}a^{3}H}\quad\quad \Omega \equiv \frac{\sqrt{8\pi G}}{\sqrt{3}H}\sqrt{\rho} e^{\phi}
\end{align}
Substituting the dynamical variables into the constraint equation \eqref{eq:constraint} yields the following form that defines the phase space of the hyperboloid of two sheets,
\begin{align}
    \label{eq:constraint2}&\Omega^{2} + y^{2} + 2x - \frac{2}{3}x^{2} = 1
\end{align}

The autonomous equation system can be constructed by calculating the evolution of the dynamical variables in the phase space by taking their derivative with respect to $dN\equiv Hdt$, denoted by the prime ($'$) symbol.
\begin{subequations}
\label{eq:autonomous_original}
\begin{align}
&x' = \frac{\ddot{\phi}}{H^{2}} - x\frac{\dot{H}}{H^{2}}\\
&y' = -y\left(3 + \frac{\dot{H}}{H^{2}}\right)\\
&\Omega' = \Omega\left(\frac{1}{2}\frac{\dot{\rho}}{H\rho} + x - \frac{\dot{H}}{H^{2}}\right)
\end{align}
\end{subequations}
Explicit forms of the terms in the above equations can be written in terms of the dynamical variables as follows.
\begin{subequations}
    \begin{align}
        &\frac{\dot{H}}{H^{2}} = \Omega^{2} \left(3\omega + \frac{3\lambda}{2} - 2\right) - 2x + \frac{4x^{2}}{3} + 4y^{2} - 1\\
        &\frac{\ddot{\phi}}{H^{2}} = \frac{3}{2}\left(\frac{\dot{H}}{H^{2}} + 1\right) + \frac{3\Omega^{2}}{2}\left(1 - \frac{\lambda}{2}\right)\\
        &\frac{\dot{\rho}}{H\rho} = -3(1 + \omega) - x\lambda
    \end{align}
\end{subequations}

The region of acceleration can be realized where  $\ddot{a}$ has positive values. In terms of the dynamical variables and equation of state parameters, this can be written as,
\begin{align}
    &H^{2}\left(\dfrac{\dot{H}}{H^{2}} + 1\right) = \dfrac{\ddot{a}}{a} > 0\\
    &\Omega^{2} \left(3\omega + \frac{3\lambda}{2} - 2\right) - 2x + \frac{4x^{2}}{3} + 4y^{2} > 0
\end{align}

As it has been stated in \citep{Angus2020}, since we take the equation of state and the dilaton's coupling parameter as constants, the flux can be taken as zero for which its contribution can then be incorporated into the energy density by setting specific values of $\omega$ and $\lambda$, i.e. $\left(\omega,\lambda\right)_{\text{h}} = (1,2)$. Therefore, from now on, we will set the flux contribution to zero, $h=0$, and we will investigate the contribution of the flux term by setting parameters accordingly. Moreover, by setting $h=0$ we reduced our phase space dimension to 2 with phase space only constructed with $x$ and $\Omega$.

\begin{figure}[!htp]
    \centering
    \begin{subfigure}[b]{0.45\textwidth}
        \includegraphics[width=\linewidth]{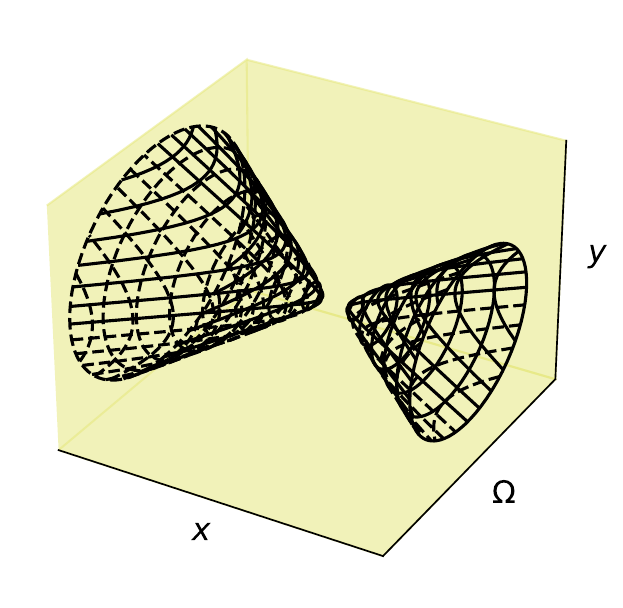}
    \caption{\label{fig:hyper3d}Hyperboloid phase space in eq.~\eqref{eq:constraint2}}
    \end{subfigure}
    \hfill
    \begin{subfigure}[b]{0.45\textwidth}
        \includegraphics[width=\linewidth]{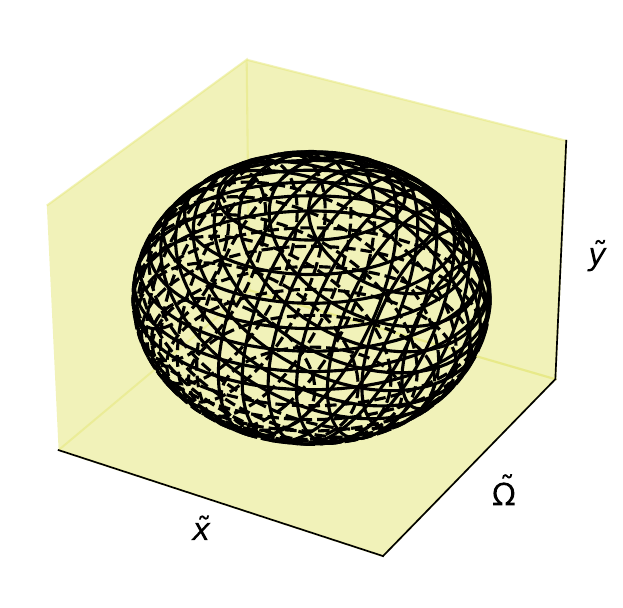}
    \caption{\label{fig:spher3d}Spherical phase space in eq.~\eqref{eq:constraint3}}
    \end{subfigure}
    \caption{\label{fig:phase_space_visuals}Visual representation of the phase spaces.}
\end{figure}

The initial phase space of the system has the hyperboloid of two sheets form as can be seen in Figure \ref{fig:hyper3d}. One can make a variable redefinition as shown below to reconstruct the phase space as a unit sphere,
\begin{align}
    \label{eq:variables2}\tilde{x} \equiv \frac{\sqrt{3}}{2\left(x-3/2\right)}\quad\quad \tilde{y} \equiv \frac{\sqrt{3}y}{\sqrt{2}\left(x-3/2\right)}\quad\quad \tilde{\Omega} \equiv \frac{\sqrt{3}\Omega}{\sqrt{2}\left(x-3/2\right)}
\end{align}
for which the phase space has now taken the form as in Figure \ref{fig:spher3d} and the constraint equation now reads,
\begin{align}
    \label{eq:constraint3}\tilde{x}^{2} + \tilde{y}^{2} + \tilde{\Omega}^{2} = 1.
\end{align}

The evolution of the phase space variables can be found as before while taking into account the evolution of the old set in terms of the new variable set. It takes a similar form as before with the addition of new terms as,
\begin{subequations}
\label{eq:autonomous_new}
\begin{align}
    \label{eq:an_x}&\tilde{x}' = -\frac{2\tilde{x}^{2}}{\sqrt{3}}\left(\frac{\ddot{\phi}}{H^{2}} - \left(\frac{3}{2} + \frac{\sqrt{3}}{2\tilde{x}}\right)\frac{\dot{H}}{H^{2}}\right),\\
    \label{eq:an_y}&\tilde{y}' = -\tilde{y}\left(3 + \frac{\dot{H}}{H^{2}}\right) - \frac{2\tilde{x}\tilde{y}}{\sqrt{3}}\left(\frac{\ddot{\phi}}{H^{2}} - \left(\frac{3}{2} + \frac{\sqrt{3}}{2\tilde{x}}\right)\frac{\dot{H}}{H^{2}}\right),\\
    \label{eq:an_om}&\tilde{\Omega}' = \tilde{\Omega}\left(\frac{1}{2}\frac{\dot{\rho}}{H\rho} + \frac{3}{2} + \frac{\sqrt{3}}{2\tilde{x}} - \frac{\dot{H}}{H^{2}} - \frac{2\sqrt{3}}{3}\tilde{x}\left(\frac{\ddot{\phi}}{H^{2}} - \left(\frac{3}{2} + \frac{\sqrt{3}}{2\tilde{x}}\right)\frac{\dot{H}}{H^{2}}\right)\right).
\end{align}    
\end{subequations}
Again we kept some of the terms closed to improve the readability of the autonomous equation set and provide their explicit form below.
\begin{subequations}
    \begin{align}
        &\frac{\dot{H}}{H^{2}} = \frac{1}{2}\frac{\tilde{\Omega}^{2}}{\tilde{x}^{2}}\left(3\omega + \frac{3\lambda}{2} - 2\right) + 2\frac{\tilde{y}^{2}}{\tilde{x}^{2}} + \frac{1}{\tilde{x}^{2}} + \frac{\sqrt{3}}{\tilde{x}} - 1\\
        &\frac{\ddot{\phi}}{H^{2}} = \frac{3}{2}\left(\frac{\dot{H}}{H^{2}} + 1\right) + \frac{3\tilde{\Omega}^{2}}{4\tilde{x}^{2}}\left(1 - \frac{\lambda}{2}\right)\\
        &\frac{\dot{\rho}}{H\rho} = -3(1 + \omega) - \left(\frac{3}{2} + \frac{\sqrt{3}}{2\tilde{x}}\right)\lambda
    \end{align}
\end{subequations}
The acceleration region of the phase space can be written as before with $\frac{\dot{H}}{H} + 1 > 0$ condition,
\begin{align}
        \label{eq:acceleration_region}&\frac{1}{2}\frac{\tilde{\Omega}^{2}}{\tilde{x}^{2}}\left(3\omega + \frac{3\lambda}{2} - 2\right) + 2\frac{\tilde{y}^{2}}{\tilde{x}^{2}} + \frac{1}{\tilde{x}^{2}} + \frac{\sqrt{3}}{\tilde{x}} > 0
\end{align}
Ultimately, both phase spaces have the same physics, for which one critical point found in one phase space can easily be converted to the critical point found in the other phase space by utilizing the definitions of the dynamical variables. Moreover, the discontinuity in the hyperboloid of two-sheet space is also present in the spherical phase space. Even though the spherical phase space looks closed, the conversion between them has a discontinuity, thus the original behavior of the system remains unchanged. Therefore, in the rest of the paper, we will use the original equation set in eq.~\eqref{eq:autonomous_original} to determine the critical points of the phase space and their characteristics.

\section{Critical Points and Stability Analysis}
\begin{figure}[htp]
    \centering
    \begin{subfigure}[b]{0.3\textwidth}
        \includegraphics[width=\linewidth]{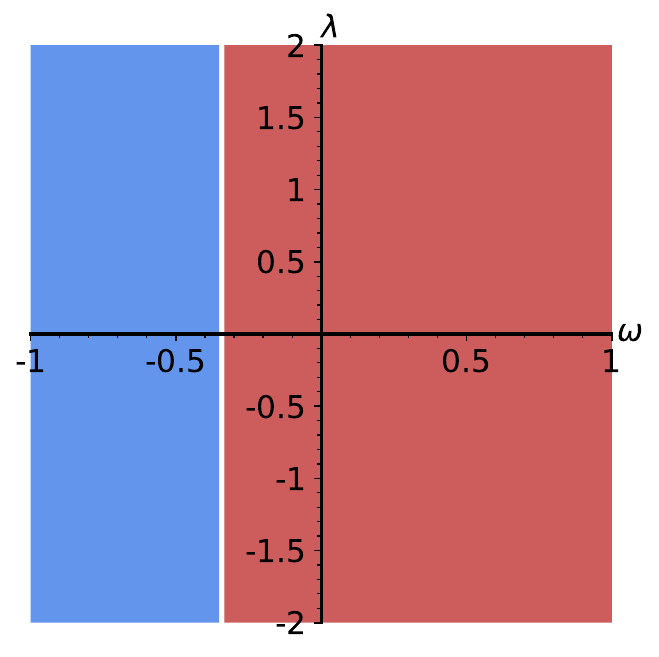}
    \caption{\label{fig:olA}Point A}
    \end{subfigure}
    \begin{subfigure}[b]{0.3\textwidth}
        \includegraphics[width=\linewidth]{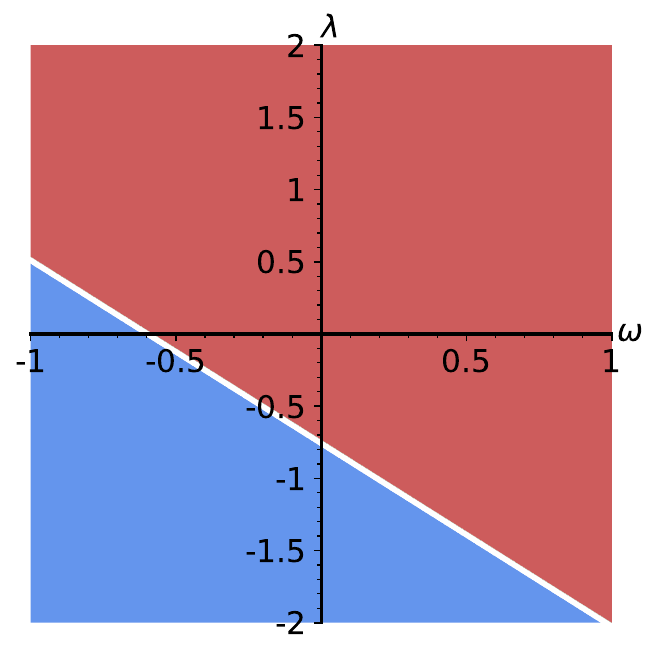}
    \caption{\label{fig:olB}Point B}
    \end{subfigure}
    \begin{subfigure}[b]{0.3\textwidth}
        \includegraphics[width=\linewidth]{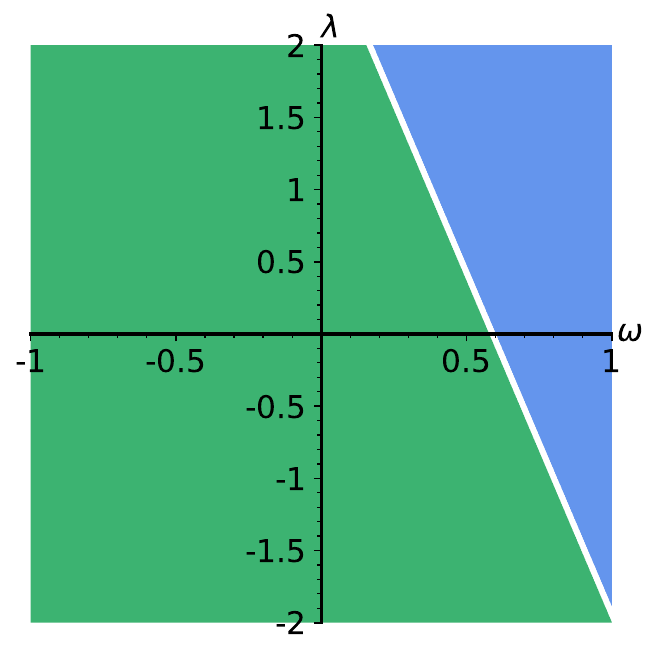}
    \caption{\label{fig:olC}Point C}
    \end{subfigure}
    \begin{subfigure}[b]{0.3\textwidth}
        \includegraphics[width=\linewidth]{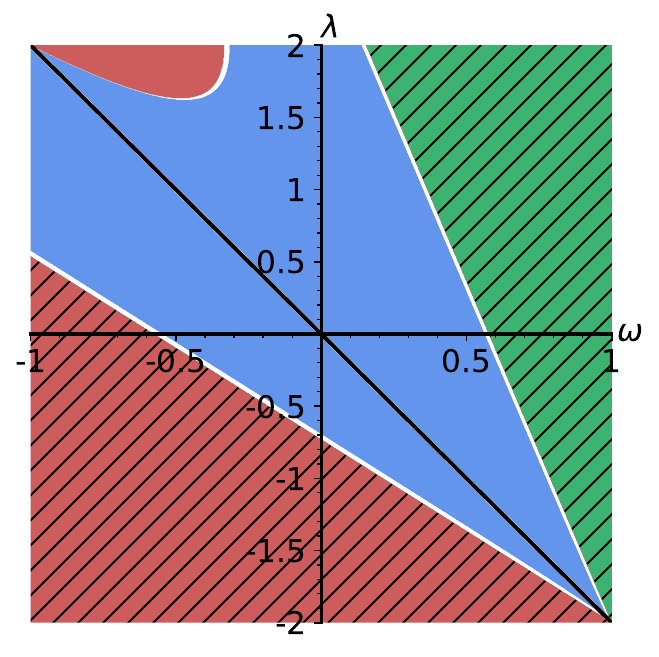}
    \caption{\label{fig:olD}Point D$_\pm$}
    \end{subfigure}
    \begin{subfigure}[b]{0.3\textwidth}
        \includegraphics[width=\linewidth]{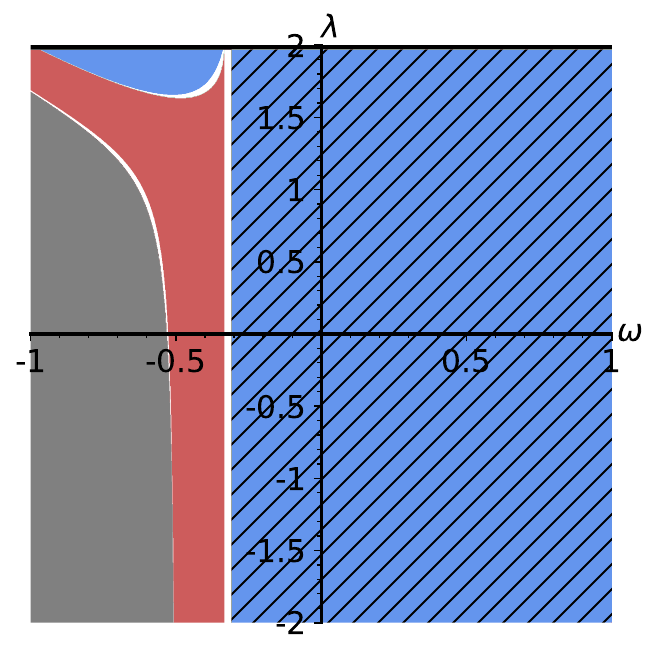}
    \caption{\label{fig:olE}Point E$_\pm$}
    \end{subfigure}
    \caption{\label{fig:bifurcation_o2}Critical point behavior analysis of Table \ref{tab:x-Omega-hyper-criticalpoints-o2} with respect to equation of state parameters $\omega$ and $\lambda$. White curves between regions represent bifurcation curves, changes in the thickness of the lines are purely for the enhancement of the visual representation. Green: Source node, Blue: Saddle node, Red: Attractor node, Grey: Stable spiral,  Black line ($\lambda = -2\omega$, $\lambda = -2$) and the shaded regions ($h(\omega,\lambda) < 0$, $\omega > -1/3$) represents the forbidden regions.}
\end{figure}

The Jacobian matrix of autonomous equation system in eq.~\eqref{eq:autonomous_original} (reduced to 2 dimensions by setting flux to zero ($h=0$)) can be found as following,
\begin{align}
\mathcal{J} = \begin{pmatrix}
-2(2x^{2} - 4x + 1) - 3\Omega^{2}(\omega + \frac{\lambda}{2} - \frac{2}{3}) & -\Omega(3\lambda(x-1) + 3\omega(2x - 3) - 4x + 3)\\
-\frac{\Omega}{6}(3\lambda + 16x - 18) & -\frac{1}{2}(x\lambda + 3w + 1) - \frac{3\Omega^{2}}{2}(6\omega + 3\lambda - 4) + 3x - \frac{4x^{2}}{3}
\end{pmatrix},
\end{align}
and the system has a total of $7$ critical points listed in Table \ref{tab:x-Omega-hyper-criticalpoints-o2} along with the critical point's existence requirement, the condition for the characteristics of the critical point and the point's behavior. We also give the corresponding bifurcation diagrams in Figure \ref{fig:bifurcation_o2} to visualize the characteristics of the critical points of the autonomous equation system based on the values of $\omega$ and $\lambda$.

\bgroup
\def\arraystretch{1.2}%
\begin{table}[htp]
\begin{tabular}{c | c | c | l | l | l | l}
\hline
\footnotesize{\textbf{\#}} & \footnotesize{\boldmath$x$} & \footnotesize{\boldmath{$\Omega$}} & \footnotesize{\textbf{Eigenvalues}} & \footnotesize{\textbf{Existence}} & \footnotesize{\textbf{Condition}} & \footnotesize{\textbf{Behavior}}\\
\hline
\hline
\hspace{-0.0cm}\small{A}\hspace{-0.0cm}   & $\hspace{-0.0cm}0\hspace{-0.0cm}$ & $\hspace{-0.0cm}0\hspace{-0.0cm}$      & $\small{\begin{array}{ll}&\hspace{-0.0cm}\lambda_{1} = -2\\&\hspace{-0.0cm}\lambda_{2} = \frac{-(3\omega + 1)}{2}\end{array}}$ &\hspace{-0.0cm} $\footnotesize{\begin{array}{ll}&\forall(\omega,\lambda)\end{array}}$ & $\footnotesize{\begin{array}{ll}&\hspace{-0.0cm}\omega < -1/3\\&\hspace{-0.0cm}\omega > -1/3\end{array}}$ & $\footnotesize{\begin{array}{ll}&\hspace{-0.0cm}\text{saddle}\\&\hspace{-0.0cm}\text{attractor}\end{array}}$\\
\hline
\hspace{-0.0cm}\small{B}\hspace{-0.0cm}   & $\hspace{-0.0cm}\frac{3+\sqrt{3}}{2}\hspace{-0.0cm}$ & $\hspace{-0.0cm}0\hspace{-0.0cm}$      & $\small{\begin{array}{ll}&\hspace{-0.0cm}\lambda_{1} = -2 - 2\sqrt{3}\\&\hspace{-0.0cm}\lambda_{2} = \frac{-\lambda(3 + \sqrt{3}) - 2(3w + \sqrt{3})}{4}\end{array}}$ &\hspace{-0.0cm} $\footnotesize{\begin{array}{ll}&\forall\left(\omega,\,\lambda\right)\end{array}}$ & $\footnotesize{\begin{array}{ll}&\hspace{-0.0cm}\lambda_{2} > 0\\&\hspace{-0.0cm}\lambda_{2} < 0\end{array}}$ & $\footnotesize{\begin{array}{ll}&\hspace{-0.0cm}\text{saddle}\\&\hspace{-0.0cm}\text{attractor}\end{array}}$\\
\hline
\hspace{-0.0cm}\small{C}\hspace{-0.0cm}   & $\hspace{-0.0cm}\frac{3-\sqrt{3}}{2}\hspace{-0.0cm}$ & $\hspace{-0.0cm}0\hspace{-0.0cm}$      & $\small{\begin{array}{ll}&\hspace{-0.0cm}\lambda_{1} = -2 + 2\sqrt{3}\\&\hspace{-0.0cm}\lambda_{2} = \frac{-\lambda(3-\sqrt{3}) - 2(3w - \sqrt{3})}{4}\end{array}}$ & \hspace{-0.0cm}$\footnotesize{\begin{array}{ll}&\forall\left(\omega,\,\lambda\right)\end{array}}$ & $\footnotesize{\begin{array}{ll}&\hspace{-0.0cm}\lambda_{2} > 0\\&\hspace{-0.0cm}\lambda_{2} < 0\end{array}}$ & $\footnotesize{\begin{array}{ll}&\hspace{-0.0cm}\text{source}\\&\hspace{-0.0cm}\text{saddle}\end{array}}$\\
\hline
\hspace{-0.0cm}\small{D$_\pm$}\hspace{-0.0cm}   & $\hspace{-0.0cm}\frac{g(\omega,\lambda)}{\lambda + 2\omega}\hspace{-0.0cm}$ & $\hspace{-0.0cm}\frac{\pm\sqrt{h(\omega,\lambda)}}{\lambda + 2\omega}\hspace{-0.0cm}$         &   $\small{\begin{array}{ll}&\hspace{-0.0cm}\lambda_{1}=\frac{g(\omega,\lambda)^{2} - 3(\omega - 1)^{2}}{2(\lambda + 2\omega)}\\[5pt]&\hspace{-0.0cm}\lambda_{2}= 2\lambda_{1} - \frac{4(\omega - 1)}{\lambda + 2\omega}\end{array}}$ & \hspace{-0.0cm}$\footnotesize{\begin{array}{ll}&\lambda \ne -2\omega\\&h(\omega,\lambda) \ge 0 \end{array}}$ &  $\footnotesize{\begin{array}{ll}&\hspace{-0.0cm}\lambda_{1,2}>0\\&\hspace{-0.0cm}\lambda_{1}\cdot\lambda_{2}<0\\&\hspace{-0.0cm}\lambda_{1,2}<0\end{array}}$ & $\footnotesize{\begin{array}{ll}&\hspace{-0.0cm}\text{source}\\&\hspace{-0.0cm}\text{saddle}\\&\hspace{-0.0cm}\text{attractor}\end{array}}$\\
\hline
\hspace{-0.0cm}\small{E$_\pm$}\hspace{-0.0cm}   & $\hspace{-0.0cm}\frac{-(3\omega + 1)}{\lambda - 2}\hspace{-0.0cm}$ & \small{$\hspace{-0.0cm}\pm\sqrt{\frac{-4(3\omega + 1)}{3(\lambda-2)^{2}}}\hspace{-0.0cm}$}     & $\small{\begin{array}{ll}&\hspace{-0.0cm}\lambda_{1} = \frac{g(\omega,\lambda) + f(\omega,\lambda)}{2-\lambda}\\&\hspace{-0.0cm}\lambda_{2} = \frac{g(\omega,\lambda) - f(\omega,\lambda)}{2-\lambda}\end{array}}$ &\hspace{-0.0cm} $\footnotesize{\begin{array}{ll}&\omega \le -1/3\\&\lambda \ne 2\end{array}}$ & $\footnotesize{\begin{array}{ll}&\hspace{-0.0cm}\lambda_{1}\cdot\lambda_{2}<0\\&\hspace{-0.0cm}\lambda_{1,2}<0\\&\hspace{-0.0cm}g<0,f^{2}<0\end{array}}$ & $\footnotesize{\begin{array}{ll}&\hspace{-0.0cm}\text{saddle}\\&\hspace{-0.0cm}\text{attractor}\\&\hspace{-0.0cm}\text{stable spiral}\end{array}}$\\
\hline
\hline
\multicolumn{7}{c}{$\small{\begin{array}{ll}&f(\omega,\lambda) \equiv \sqrt{6 \lambda^{2} \omega + 3 \lambda^{2} + 36 \lambda \omega^{2} + 6 \lambda \omega - 6 \lambda + 36 \omega^{3} - 3 \omega^{2} - 2 \omega + 5}\\&g(\omega,\lambda) \equiv \lambda + 3\omega - 1\qquad h(\omega,\lambda) = (\omega-1)^{2} - \frac{1}{3}g(\omega,\lambda)^{2}\end{array}}$}\\
\hline
\hline
\end{tabular}
\caption{\label{tab:x-Omega-hyper-criticalpoints-o2}Critical points of autonomous system in eq.~\eqref{eq:autonomous_original}}
\end{table}
\egroup

\begin{itemize}
\item \textbf{Point A} always exists without any restrictions coming from the parameters $\omega$ and $\lambda$. It lies at the origin of the phase space as $(x,\Omega) = (0,0)$ with the behavior depending solely on the parameter $\omega$. In the case of $\omega < -1/3$ the point behaves as a saddle point, whereas when the $\omega$ passes the value $-1/3$ the point behaves as an attractor node. The behavior at the intersection of those two regions where the $\omega$ takes the value exactly $-1/3$, which characterizes the curvature contribution, can not be determined by the linear stability analysis since the point then becomes non-hyperbolic as it will have a zero eigenvalue. As can also be seen in the Figure \ref{fig:olA}, for the majority of the allowed values of $(\omega,\lambda)$ pair, the critical point appears as an attractor node, in which the point corresponds to the complete flux contribution universe, i.e. $y = 1$.

\item \textbf{Point B} is independent of the other critical points, meaning that the field lines that live in the sheet where critical point B exists, never cross to the other sheet that houses the remaining critical points because of the hyperboloid of two sheet characteristics which can be seen in Figure \ref{fig:hyper3d}. Moreover, the point B always exists without any dependence on the parameters $\omega$ or $\lambda$. One needs to choose phase space sheets from the existing two to continue the analysis, for which we considered the sheet that constrains the $x$ as $x < 3/2$. Justification of this particular choice can be realized when we think about the constraint equation in \eqref{eq:constraint2}. The $x$ contribution to the constraint equation is $2x - 2x^{2}/3$ and each squared contribution that is in the constraint equation is constrained by $0$ from the below and by $1$ from the above, i.e. for $\epsilon \in \{2x - 2x^{2}/3,\,\Omega^{2},\,y^{2}\}, 0 \le \epsilon \le 1$ demanded by the constraint. Therefore, the value of $x \ge 3/2$ is not in the scope. Thus, we will not make any further analysis on critical point B, since it is not in the scope of the phase space we have chosen.

\item \textbf{Point C} also exists without any dependence on the values of $\omega$ and $\lambda$. It can have the characteristics of both a source node and a saddle node for specific values of $\omega$ and $\lambda$. For the majority of the values of $\omega$ and $\lambda$, the point acts as a source node. The dominant contribution to the evolution at the critical point C comes from the $x = \frac{\dot{\phi}}{H}$ with $\Omega = 0$. The only possible source contribution in our phase space can be realized in the critical point C and, for the case of point A being the attractor node and point C being the source node the restriction to the equation of state parameters can be written as follows,
\begin{align}
    -\dfrac{1}{3} < \omega < \dfrac{1}{\sqrt{3}} - \dfrac{\lambda}{2}\left(1 - \dfrac{1}{\sqrt{3}}\right)
\end{align}
For which, the upper bound on $\omega$ given above can be visually understood as in the Figure \ref{fig:olC}. Moreover, since there is no restriction on the range of allowed values of the parameter $\lambda$ from $-2$ to $2$, the above inequality has the upper bound stated by the parameter $\lambda$ as,
\begin{align}
    -\dfrac{1}{3} < \omega < \begin{cases}\quad 1 \ \ \, \quad \rightarrow \lambda = -2\\\frac{2\sqrt{3}}{3} - 1 \rightarrow \lambda = 2\end{cases}
\end{align}
In the case of $\lambda = 2$ the upper bound becomes $\omega \lesssim 0.15$, for which the choice of $\omega = 1$ for the flux contribution integrated into the $\rho$, $(\omega,\lambda)_{\rm{h}} = (1, 2)$, results in with a phase space that does not have a source node.

\item \textbf{Point D$_\pm$} points can not exist simultaneously, because of the $\Omega > 0$ condition. In the positive case of the critical point, the condition of existence is constrained by the $\lambda + 2\omega > 0$ statement, and in the negative case, the constraining condition becomes the opposite as $\lambda + 2\omega < 0$. Without taking the conditions for existence the point's stability characteristics can be categorized as; for the negative case, the condition reduces the possible values that can be taken for the $\omega,\,\lambda$ to the below of the black curve in Figure \ref{fig:olD}. For this, the only possible characteristic of the critical point becomes a saddle-node or an attractor node. For the positive case, the point can take all three characteristics as a saddle, an attractor, or a source node without the change in location of the point in phase space. Moreover, the shaded areas in the figure represent the values of $\omega$ and $\lambda$ where the $h(\omega,\lambda)$ becomes negative, which renders $\Omega$ parameter imaginary, therefore can not be taken as a possible solution and completely removes the source characteristics from the critical point.

\item \textbf{Point E$_\pm$}'s condition of existence is restricted by both parameters $\omega$ and $\lambda$ where the restricted region that is dependent on the parameter $\omega$ is given as $\omega > -1/3$, for which makes the $\Omega$ variable complex-valued as can be seen in Figure \ref{fig:olE}. On the other hand, the restriction coming from the parameter $\lambda$ is for its specific value $\lambda = 2$ which in that case both dynamical variables $x$ and $\Omega$ go to infinity. In the allowed region where $\omega \ge -1/3$ and $\lambda \ne 2$, the critical point E can be a saddle-node, an attractor node, or a stable spiral node based on the values of $\omega$ and $\lambda$ that is given in the Table \ref{tab:x-Omega-hyper-criticalpoints-o2}.

\end{itemize}

\section{Cosmologically Interesting Cases}
We investigate the phase space dynamics with different parameter sets of cosmologically interesting cases. The initial condition values for the dynamical variables $x$ and $\Omega$ are chosen such that the evolution of the effective equation of state parameter $\omega_{\rm{eff}}$ starts with the value close to that of the radiation, i.e. $\omega_{\rm{eff}} = 1/3$, where the $\omega_{\rm{eff}}$ is defined using the eq.~\eqref{eq:odd_feq1} as follows,

\begin{align}
    3H^{2} &= 8\pi G \rho_{\rm{eff}}\quad\text{and}\quad6H\dot{H} = 8\pi G (-3H\rho_{\rm{eff}}(1 + \omega_{\rm{eff}}))
\end{align}
which gives,
\begin{align}
    \omega_{\rm{eff}} &= - 1 - \dfrac{2}{3}\dfrac{\dot{H}}{H^{2}}.
\end{align}

We also provide in the figures below, the evolution of the deceleration parameter $q$ defined as,
\begin{align}
    q \equiv -1 - \dfrac{\dot{H}}{H^{2}}.
\end{align}
The deceleration parameter $q$ is useful to determine the acceleration of the universe:  the universe is accelerating when $q < 0$ and decelerating when $q > 0$. We reintroduced the dynamical variable $y$ as a derived parameter from the constraint equation eq.~\eqref{eq:constraint2}. We had previously set it to zero without loss of generality as the flux contribution can be integrated into the energy density by employing specific values for the equation of state parameters and provided its evolution along with other two dynamical variables $x$ and $\Omega$. Consequently, the flux domination, $y=1$ is the attractor of the system for the values of $\omega > -1/3$ and no restrictions on the value of $\lambda$, as it can be seen in the bifurcation figures \ref{fig:bifurcation_o2}. Apart from the cosmological constant which has the $\omega = -1$, the critical point $A$ is the attractor of the phase space. Below, we present the phase space trajectories of configurations, accompanied by the evolution of the dynamical variables $x$, $y$ and $\Omega$, the effective equation of state $\omega_{\rm{eff}}$ and the deceleration parameter $q$ for a given $w$, $\lambda$ and the initial condition for the $x$ and $\Omega$.
\subsection{Scalar field, $(\lambda = -2\omega),\, (\omega, \lambda) = (-1/3, 2/3)$}
\begin{figure}[htp]
    \centering
    \includegraphics[width=0.9\textwidth]{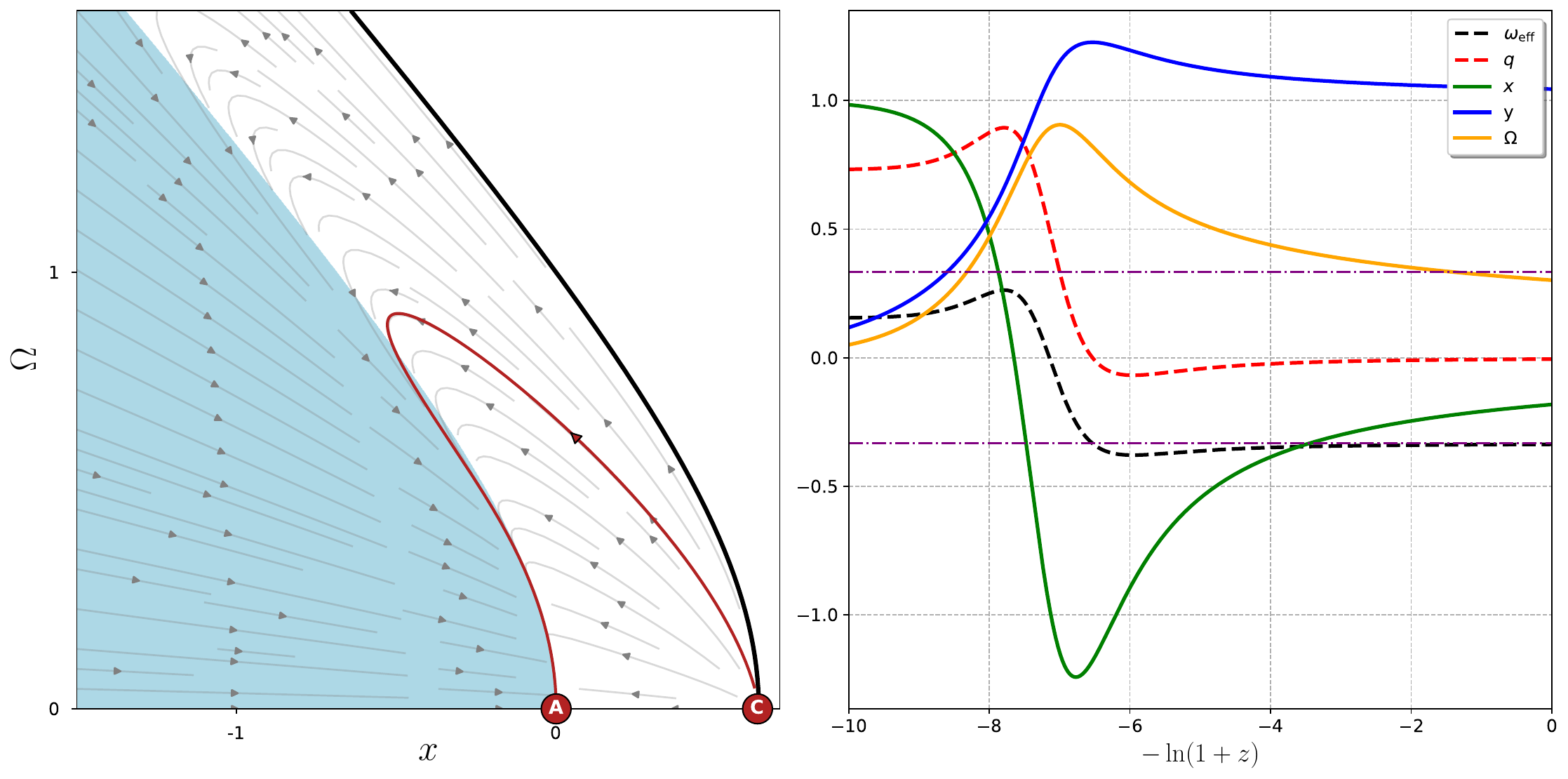}
    \caption{On the left-hand side, we give the phase space trajectory for the scalar field with initial conditions $(x,\Omega) = (0.62, 0.05)$ that corresponds to the source node C. The blue shaded area corresponds to the accelerating region of the phase space with $\frac{\dot{H}}{H^{2}} + 1 > 1$. On the right-hand side, with the same initial conditions, we give the evolution of the dynamical system variables ($\Omega$, $x$, and $y$), effective equation of state parameter ($\omega_{\rm{eff}}$) and the deceleration parameter ($q$). Horizontal dashed lines correspond to the values $\pm 1/3$.}
    \label{fig:sf-phase-space}
\end{figure}

In the case of scalar-field-like contribution to the energy density $\rho$ we take the equation of state parameters in such a way that $\lambda = -2\omega$. By choosing the value of the $\omega$ as $-1/3$ we arrive at the following phase space diagram and the parameter evolution in Figure \ref{fig:sf-phase-space}. There are only two critical points in the phase space $A$ and $C$ corresponding to the points with the same name in Table \ref{tab:x-Omega-hyper-criticalpoints-o2}. The acceleration region of the phase space can be seen with the blue shaded area on the left-hand side of the Figure \ref{fig:sf-phase-space}. The phase space trajectories start from the source node $C$ for which the dominating contribution comes from the kinetic energy of the scalar field and evolves to the attractor node $A$ for which both $x$ and $\Omega$ at that point vanishes with the only contribution coming from the flux term $y$. The evolution of the parameters of the model can be seen on the right-hand side of the above figure. One can see that the universe begins its acceleration phase when the flux contribution starts to dominate the universe. Furthermore, as the flux domination settles to unity, the effective equation of state reaches the value of $\omega_{\rm{eff}} = -1/3$.

\subsection{Radiation, $(\omega,\lambda) = (1/3, 0)$}
\begin{figure}[htp]
    \centering
    \includegraphics[width=0.9\textwidth]{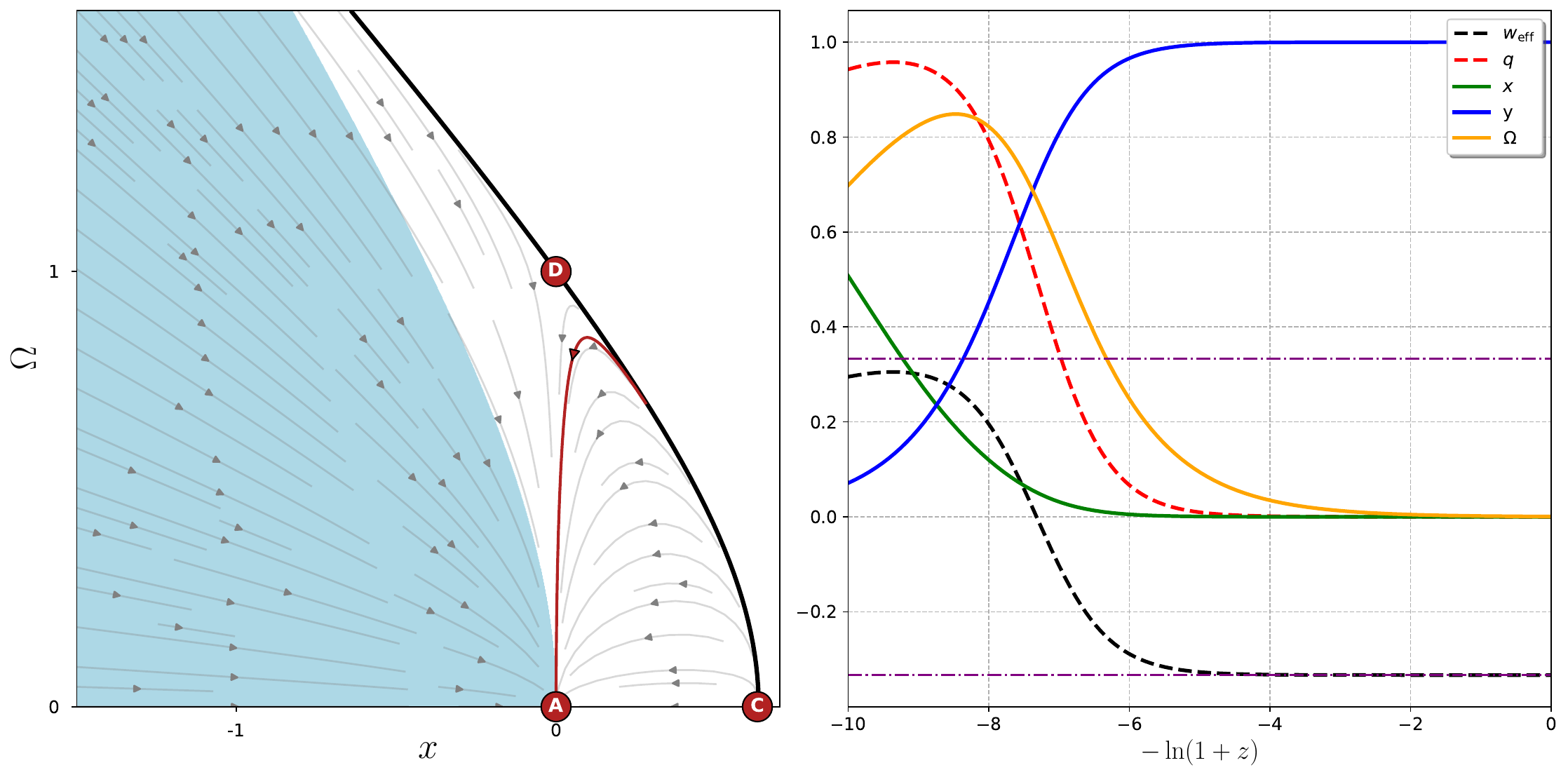}
    \caption{The phase space trajectory and parameter evolution for the radiation contribution with the initial conditions $(x,\Omega) = (0.28, 0.69)$ where again the blue shaded area corresponds to the accelerating region of the phase space with $\frac{\dot{H}}{H^{2}} + 1 > 1$. Horizontal dashed lines correspond to the values $\pm 1/3$.}
    \label{fig:rd-phase-space}
\end{figure}

In the configuration for a radiation contribution, we have a total of three critical points in the phase space as can be seen on the left-hand-side of the Figure \ref{fig:rd-phase-space}. Taking the initial conditions for the variables $x$ and $\Omega$ as $(0.28, 0.69)$, the field line that shows the evolution of the phase space in this configuration starts from the source point $C$ and follows a path towards the saddle-node $D$ and eventually the trajectory resolves to the attractor-node $A$. The dominant contributions for each node in the phase space can be understood by looking at the right-hand side of the same figure, for which the universe now starts in the radiation-dominated phase with the major contribution coming from the variable $\Omega$. In this starting configuration, the initial effective equation of state has the value of $\omega_{\rm{eff}} \simeq 1/3$, while the deceleration parameter starts close to unity, $q \simeq 1$. As the trajectory approaches the saddle node, the flux contribution starts to dominate and eventually reaches unity, i.e. $y = 1$ when the trajectory arrives at the attractor-node $A$. The effective equation of state settles on the value $w_{\rm{eff}} = -1/3$ and the deceleration parameter reaches $0$ but does not go beyond that point for which the universe accelerates, $q \nless 0$. In other words, the configuration does not evolve to a universe where it accelerates. During the evolution, the kinetic energy term of the scalar field relaxes to the value of $0$.

\subsection{Dust, $(\omega, \lambda) = (0, 0)$}
\begin{figure}[htp]
    \centering
    \includegraphics[width=0.9\textwidth]{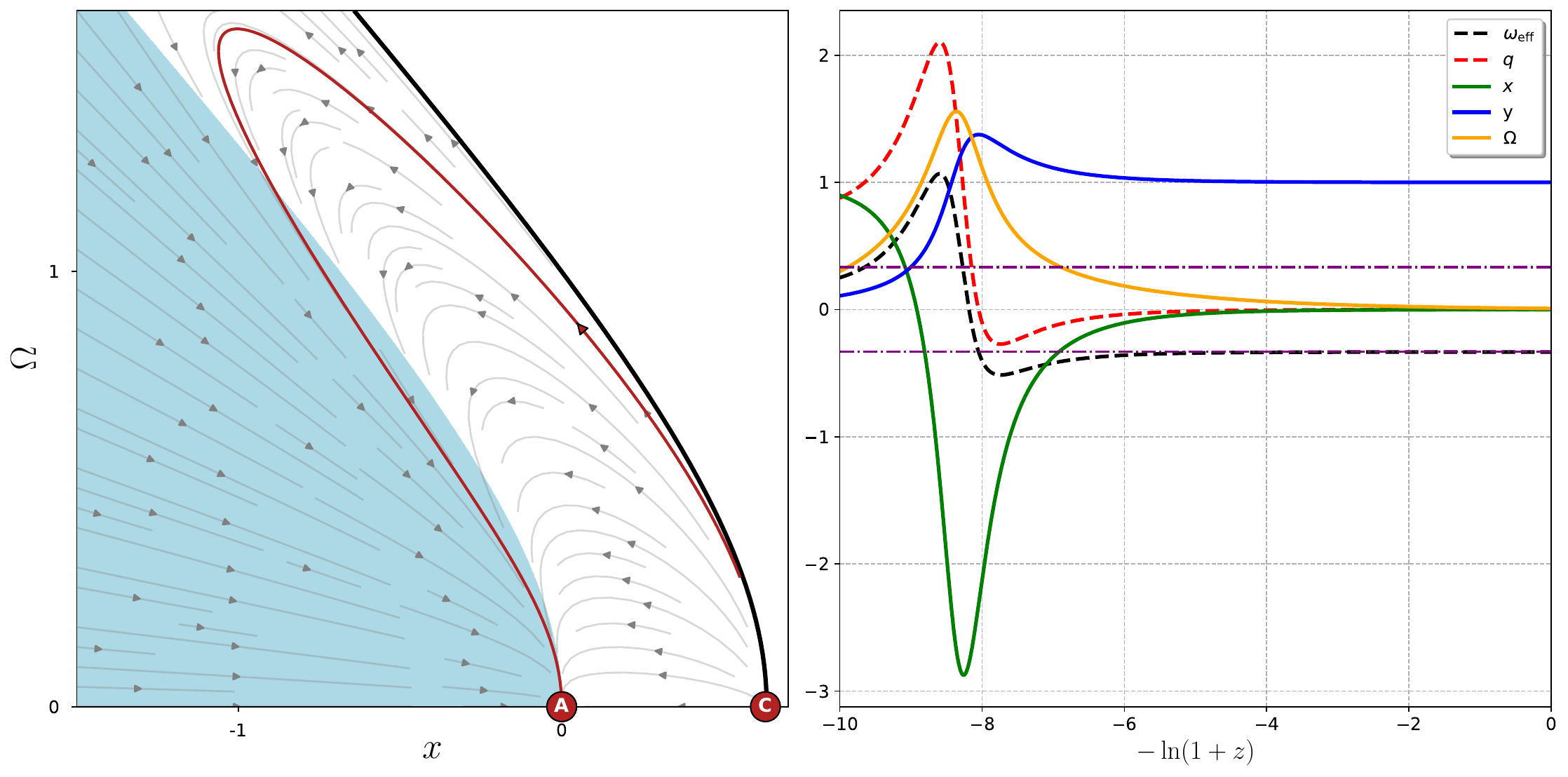}
    \caption{The phase space trajectory and parameter evolution for dust contribution with the initial values are taken as $(x, \Omega) = (0.55, 0.30)$. The black dashed line corresponds to the effective equation of state evolution, whereas the red dashed line corresponds to the deceleration parameter evolution.}
    \label{fig:dust-phase-space}
\end{figure}

In the case of a dust contribution, the equation of state parameters takes the value $(\omega, \lambda) = (0, 0)$ for which it also coincides with the scalar field contribution since in general it is given as $\lambda = -2\omega$. In Figure \ref{fig:dust-phase-space}, we present the phase space diagram for the dust configuration and parameter evolution for the initial condition $(x, \Omega) = (0.55, 0.30)$. Similar to the scalar field case, phase space has two critical points, $A$ and $C$ being attractor and source nodes respectively. The initial value for the effective equation of state, $\omega_{\rm{eff}}$, has a value close to $1/3$, and after evolution, it settles on the value $-1/3$. The evolution graph and the trajectory show that the acceleration again starts as the flux contribution starts to dominate the universe. The kinetic energy of the scalar field contribution, $x$, rapidly decreases as the universe evolves until the flux domination, after which it starts to grow and eventually resolves on the zero.

\section{Chameleon Cosmology}
The chameleon mechanism is a scenario that, in accordance with string theory predictions, permits scalar fields to evolve on current cosmological time scales with couplings of order unity to matter. This mechanism is considered as one of the so-called `screening' mechanisms\footnote{A closely related one is the Damour-Polyakov mechanism and the other is Vainstein mechanism.} in which the mass of the scalar field depends on the density of local matter: the field can be sufficiently massive to fulfill limits on equivalence principle violations and fifth force in high-density areas like the solar system but in cosmological scales evolve with a mass of the order of the Hubble parameter. The screened dynamics of the scalar field in the chameleon mechanism is the result of the interplay between a non-linear potential and a linear coupling to matter. In a symmetry-dictated, and thus more physically motivated setting like $O(D, D)$-symmetric formulation, the coupling of the dilaton to matter sector is determined on firm ground and anticipated to include the chameleon-like effects. In this section, we will elaborate on this issue.  

We have the action for the Chameleon cosmology as shown below with the function $f\left(\phi\right)$ being the coupling to the matter sector,
\begin{align}
\label{eq:cham_action}
    S = \int d^{4}x \sqrt{-g}\left\{\dfrac{R}{2} - \dfrac{1}{2}g^{\mu\nu}\nabla_{\mu}\phi\nabla_{\nu}\phi - V\left(\phi\right) +f(\phi)L_{\text{m}}\right\}\quad L_{\text{m}}\sim \rho_{\text{m}}
\end{align}

Varying the action in eq.~\eqref{eq:cham_action} with respect to the scalar field $\phi$ and inverse metric we arrive at the following equations of motion,
\begin{align}
    \label{eq:cham_var_phi}&\ddot{\phi} + 3H\dot{\phi} + V'(\phi) - f'(\phi)\rho_{\text{m}} = 0\\
    \label{eq:cham_var_im}&3H^{2} = \dfrac{\dot{\phi}^{2}}{2} + V + f(\phi)\rho_{\text{m}}
\end{align}
Furthermore, by taking the derivative of the eq.~\eqref{eq:cham_var_im} and using the equation of motion for the scalar field in eq.~\eqref{eq:cham_var_phi}, we have the following expression,
\begin{align}
    &2\dot{H} + 3H^{2} = -\left(\dfrac{\dot{\phi}^{2}}{2} - V(\phi) + f(\phi)p_{\text{m}}\right)
\end{align}
The final expression we will utilize in the linear stability analysis of the Chameleon cosmology is the conservation equation which can be written as follows,
\begin{align}
    &\dot{\rho}_{\text{m}} + 3H(\rho_{\text{m}} + p_{\text{m}}) + 2\dfrac{f'}{f}\dot{\phi}\rho_{\text{m}} = 0
\end{align}
As previously done, we will construct the constraint equation for the phase space by dividing both sides of the eq.~\eqref{eq:cham_var_im} by $3H^{2}$ and define the phase space dynamical variables accordingly.
\begin{align}
    \label{eq:cham_constraint0}1 = \dfrac{\dot{\phi}^{2}}{6H^{2}} + \dfrac{V}{3H^{2}} + \dfrac{f\rho_{\text{m}}}{3H^{2}}
\end{align}
Thus, we have the following dynamical variables for the phase space of Chameleon cosmology,
\begin{align}
    \label{eq:cham_variables}&x\equiv \dfrac{\dot{\phi}}{\sqrt{6}H}\qquad y\equiv \dfrac{\sqrt{V}}{\sqrt{3}H}\qquad \Omega \equiv \dfrac{f\rho_{\text{m}}}{3H^{2}}, \quad f(\phi) \geq 0,\ \Omega \geq 0
\end{align}
Therefore, the constraint equation in eq.~\eqref{eq:cham_constraint0} can be rewritten using the dynamical variables of the phase space of the Chameleon setting as,
\begin{align}
    \label{eq:cham_constratint}&1 = x^{2} + y^{2} + \Omega
\end{align}
One can immediately see that the dynamical variable $x$ again gives the scalar field's kinetic energy contribution to the evolution with an overall constant factor. Moreover, the variable $\Omega$ in the Chameleon setting in eq.~\eqref{eq:cham_variables} resembles the one in $O(D, D)$ configuration defined with the same name in eq.~\eqref{eq:variables1}. Lastly, the variable $y$ in the Chameleon setting provides the contribution of the scalar field's potential, yet in the $O(D, D)$ setting it shows the flux contribution, $h$. 

The autonomous equation system of the Chameleon cosmology can now be found by again taking the time derivative of the dynamical variables and dividing it by $H$ which we denote it as a prime, $'$, followed by rewriting them in terms of the dynamical variables of the phase space,
\begin{subequations}
    \label{eq:cham_system}
    \begin{align}
        &x' = \dfrac{1}{\sqrt{6}}\dfrac{\ddot{\phi}}{H^{2}} - x\dfrac{\dot{H}}{H^{2}}\\
        &y' = y\left(\sqrt{\dfrac{3}{2}}\dfrac{V'}{V}x - \dfrac{\dot{H}}{H^{2}}\right)\\
        &\Omega' = -\Omega\left(\sqrt{6}\dfrac{f'}{f}x + 3(1+w) + 2\dfrac{\dot{H}}{H^{2}}\right).
    \end{align}
\end{subequations}
Explicit forms of the acceleration term of the scalar field and the derivative of the Hubble constant $H$ are given as follows,
\begin{align}
    &\dfrac{\dot{H}}{H^{2}} = -3\left(x^{2}+\dfrac{1}{2}(1+w)\Omega\right)\\
    &\dfrac{\ddot{\phi}}{H^{2}} = -3\left(\sqrt{6}x + \dfrac{V'}{V}y^{2} - \dfrac{f'}{f}\Omega\right)
\end{align}
We now choose  the potential $V(\phi)$ and the coupling $f(\phi)$ in exponential form as follows:
\begin{align}
    &V(\phi) \propto e^{A\phi},\qquad f(\phi) \propto e^{B\phi},
\end{align}
where $A$ and $B$ are constants. With this choice, the systems of equations for the  $O(D, D)$ extended system and the Chameleon cases are manifestly comparable.  One can see that the coupling term $f\left(\phi\right) \propto e^{B\phi}$ is closely related to the parameter $\lambda$ in the $O(D, D)$ complete formulation since both measure the scalar field's coupling to the matter sector. Furthermore, the scalar potential term, $y$, in the dynamical system formulation of the Chameleon cosmology eq.~\eqref{eq:cham_variables} corresponds to the flux term, $\propto h$ in eq.~\eqref{eq:variables1}. Thus, to compare the Chameleon cosmology and the $O(D, D)$ complete cosmology in the setting of linear stability analysis one can set $A$ and $B$ constants and the equation of state parameter $\omega$ accordingly, so that the contribution of each term in Chameleon cosmological setting corresponds to the ones in the previously examined cosmological cases in the $O(D, D)$ cosmology.

In our $O(D, D)$-complete cosmology linear stability analysis, we set the flux term to zero without loss of generality. Therefore, to match the contribution in the Chameleon formulation we set $A$ to be zero to have constant scalar field potential. Furthermore, by direct comparison of the $\Omega$ parameters in eq.~\eqref{eq:variables1} and eq.~\eqref{eq:cham_variables} we set $B = 2$. These choices for $A$ and $B$ also ensure that the $O(D, D)$ symmetry is not broken.\footnote{In \citep{Choi:2022srv}, the form of the Eq.(6) emphasizes the similarity to the Chameleon interpretation, if one considers the terms in the parentheses as the effective mass.}

We performed the same linear stability analysis to the corresponding Chameleon cosmology counterparts for all three previously studied cases of $O(D, D)$ cosmologies. The dynamical system phase space in eq.~\eqref{eq:cham_system} has three distinct critical points with source characteristics at the node $\left(x,\, \Omega\right) = (-1,\,0)$, with saddle characteristics at $\left(x,\, \Omega\right) = (1,\,0)$, and with the attractor node at the $\left(x,\, \Omega\right) = (0,\,0)$ which is named $A$, $B$, and $C$ respectively.

\subsection{Scalar field, $\omega = -1/3$}
\begin{figure}[htp]
    \centering
    \includegraphics[width=0.9\textwidth]{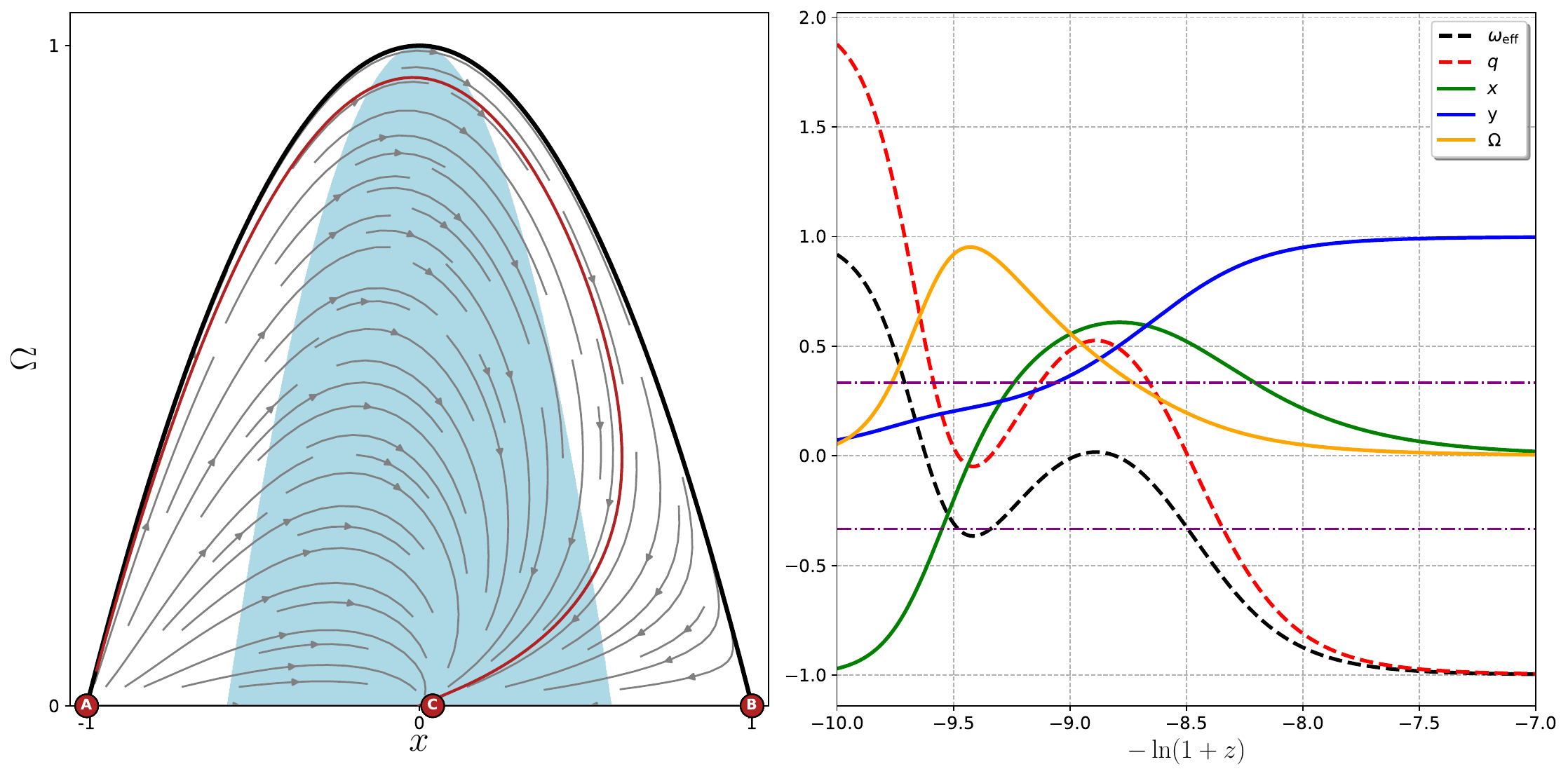}
    \caption{Phase space evolution for scalar field contribution in the Chameleon cosmology. Initial values are taken as $(x, \Omega) = (-0.97, 0.05)$.}
    \label{fig:cham-sf-phase}
\end{figure}

To study the scalar field in the context of Chameleon cosmology, we take the $\omega$ as $-1/3$, and on the left-hand side of Figure \ref{fig:cham-sf-phase} we give the resulting phase space trajectories along with on the right-hand side the phase space parameter evolution can be seen.

Phase space trajectories starts from the source node $A(x,\,\Omega) = (-1, 0)$ goes near the saddle point $B(x,\,\Omega) = (1, 0)$ and finally reaches the attractor node $C(x,\,\Omega) = (0, 0)$. The accelerating region again is a blue-shaded area where the brief acceleration occurs with $\Omega$ domination after which the second acceleration happens when the scalar field potential term $y$ comes to domination and eventually reaches the attractor node $C$. Comparison with the $O(D, D)$ symmetric setting shows that the effective equation of state $w_{\textrm{eff}}$ in the Chameleon configuration relaxes to $-1$, whereas in the $O(D, D)$ analysis the effective equation of state goes to $-1/3$ value. By looking at the deceleration parameter $q$, one can see that in the scalar field case of the Chameleon setting, there are two accelerating regions for which the first acceleration happens during the $\Omega$ domination and the second occurs when the scalar field potential term $y$ comes to dominate in the neighbor of the attractor node $C$ at the end of evolution which is a similar case in the $O(D, D)$ setting where the attractor point is the one where the flux term dominates. 
\subsection{Radiation, $\omega = 1/3$}
\begin{figure}[htp]
    \centering
    \includegraphics[width=0.9\textwidth]{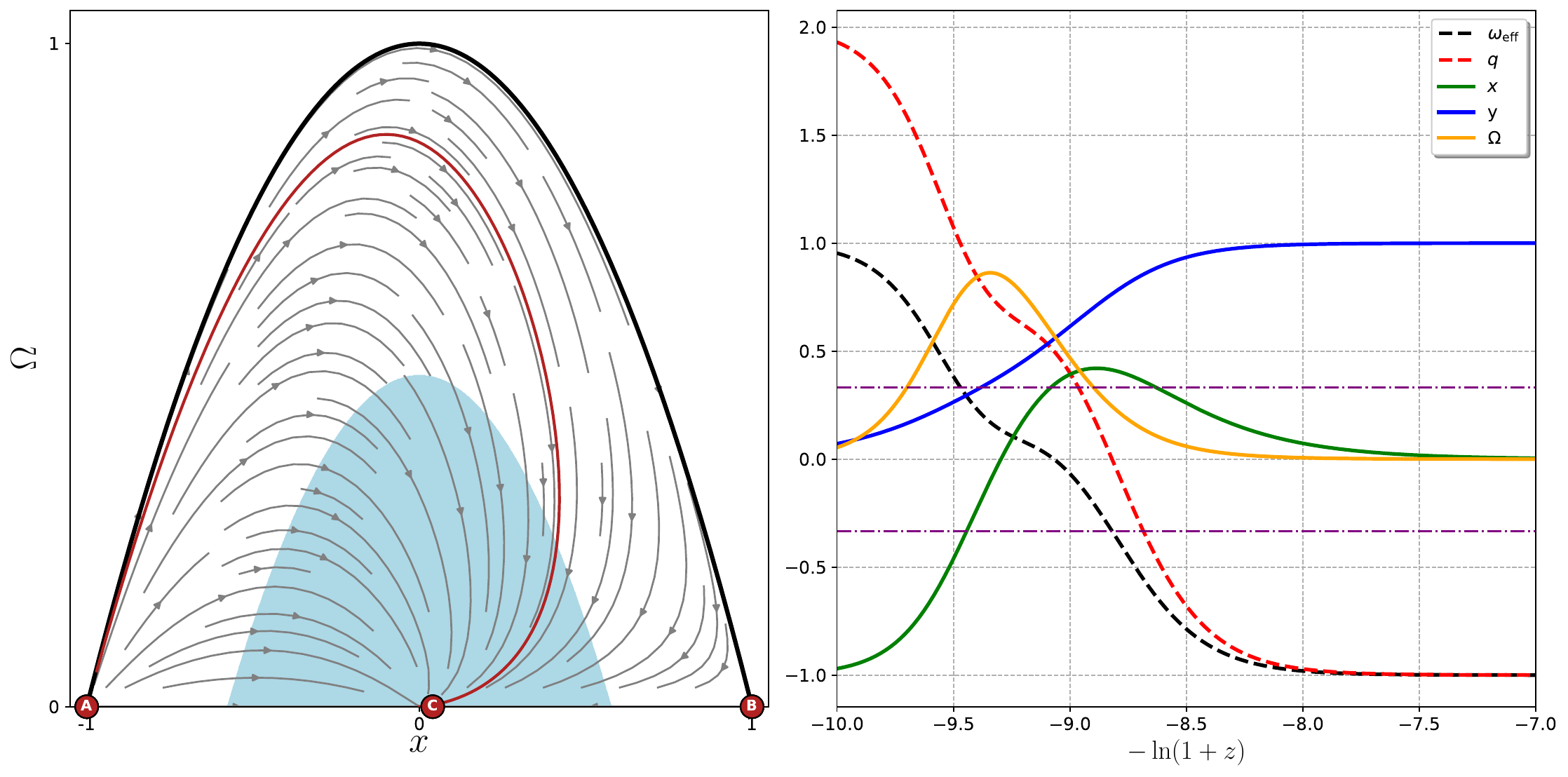}
    \caption{Phase space evolution for radiation contribution in the Chameleon cosmology. On the left-hand side, phase space trajectory with the initial condition $(x, \Omega) = (-0.97, 0.05)$ is plotted with the red line The blue shaded region represents the accelerated region and the black solid line shows the constraint. On the right-hand side, we give the parameter evolution as same as in the $O(D, D)$ symmetric setting.}
    \label{fig:cham-rad-phase}
\end{figure}

In Figure \ref{fig:cham-rad-phase} we give the phase space diagram and parameter evolution in the radiation setting. The phase space has a source node, a saddle node, and an attractor node for which we give the initial condition for the start of the trajectory to coincide with the source node as $(x,\,\Omega) = (-0.97, 0.05)$. With the given initial condition, the acceleration begins only when the scalar field potential term, $y$, becomes the dominating factor as the deceleration parameter goes below $0$ and eventually reaches $-1$, where the trajectory reaches the attractor node at $(x,\,\Omega) = (0, 0)$. The attractor region window significantly reduces in the radiation case for which the two accelerating regions in the previous case are now not apparent. Acceleration only happens when the potential of the scalar field terms becomes the dominating factor of the system. Similar to the scalar field case in the Chameleon setting, in the radiation case, the effective equation of state goes to $-1$ as opposed to $-1/3$ which was the case in the $O(D, D)$ complete cosmology analysis. The attractor point of the Chameleon system again matches to the one of a flux contribution in the $O(D, D)$ setting.

\subsection{Dust, $\omega = 0$}
\begin{figure}[htp]
    \centering
    \includegraphics[width=0.9\textwidth]{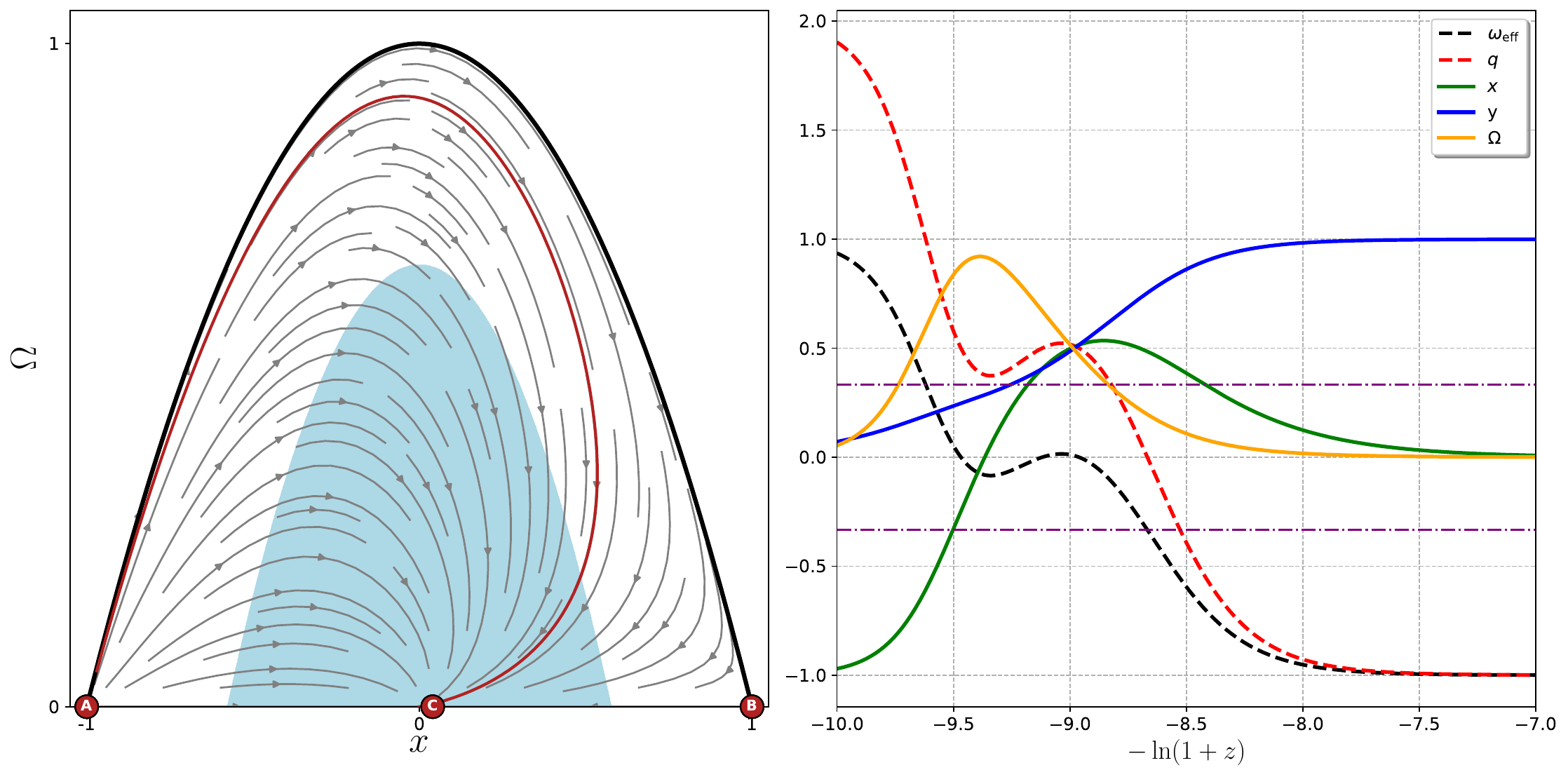}
    \caption{Phase space evolution for dust contribution in the Chameleon cosmology. Initial values are taken as $(x, \Omega) = (-0.97, 0.05)$.}
    \label{fig:cham-dust-phase}
\end{figure}

By taking the equation of state parameter as zero, we can study the contribution of dust, i.e. pressureless matter. In Figure \ref{fig:cham-dust-phase}, we give the phase space portrait of the dynamical system along with the parameter evolution. The acceleration window is narrower than the scalar field contribution yet it is wider than the radiation one. The starting point of the trajectory given in the left-hand side of the figure is taken as before, $(x,\,\Omega) = (-0.97, 0.05)$ such that it originates from the source point. The acceleration begins again when the potential term dominates which is the case for all analyzed systems for both $O(D, D)$ complete cosmology setting and the Chameleon cosmology setting. The equation of state parameter $w_{\textrm{eff}}$ again goes to the one of a scalar field or a cosmological constant $-1$, while the potential term slowly takes over the domination.

In all three cases that we have analyzed, similar aspects of $O(D, D)$ complete cosmology and the Chameleon cosmology show themselves in the eventual dominating factors of the system, which are the flux contribution and the potential of the scalar field, respectively. The striking difference is that the final value for the effective equation of state value reaches to the one of a scalar field or a cosmological constant $-1$ in the Chameleon setting, whereas in the $O(D, D)$ complete cosmology in all cases the effective equation of state goes to $-1/3$. The area of acceleration regions in the Chameleon cosmology significantly depends on the contribution of the contents of the system whereas in the $O(D, D)$ setting the difference in the accelerating regions is not that striking almost constant yet in the $O(D, D)$ symmetric setting, the acceleration window of the system very much depends on the content for which it can be seen in the case of radiation contribution of the $O(D, D)$ cosmology where the acceleration barely happens.

\section{Stability Analysis of the Case $k\ne 0$}
The $O(D, D)$-complete Friedmann equations including the spatial curvature term can be written as in the equation set \ref{eq:curved_odd_fed_set}:
\begin{subequations}
\label{eq:curved_odd_fed_set}
\begin{align}
    \label{eq:curved_odd_feq1}\frac{8\pi G}{3}\rho e^{2\phi} + \frac{h^{2}}{12 a^{6}} &= H^{2} - 2\dot{\phi}H + \frac{2}{3}\dot{\phi}^{2} + \frac{k}{a^{2}}\\
    \label{eq:curved_odd_feq2}\frac{4\pi G}{3}\left(\rho + 3p\right)e^{2\phi} + \frac{h^{2}}{6a^{6}} &= -H^{2} - \dot{H} + \dot{\phi}H - \frac{2}{3}\dot{\phi}^{2} + \ddot{\phi}\\
    \label{eq:curved_odd_feq3}\frac{8\pi G}{3}\left(\rho e^{2\phi} - \frac{1}{2}T_{(0)}\right) &= -H^{2} - \dot{H} + \frac{2}{3}\ddot{\phi}.
\end{align}
\end{subequations}

The phase space is again constructed by the same dynamical variables as in the $k=0$ case with the difference being the curvature variable denoted $\Omega_{k}$,
\begin{align}
    \label{eq:curved_variables1}x \equiv \frac{\dot{\phi}}{H}\quad\quad y \equiv \frac{h}{2\sqrt{3}a^{3}H}\quad\quad \Omega \equiv \frac{\sqrt{8\pi G}}{\sqrt{3}H}\sqrt{\rho} e^{\phi}\quad\quad \Omega_{k} = -\frac{k}{a^{2}H^{2}}.
\end{align}

The constraint equation now has the form,
\begin{align}
    \label{eq:curved_constraint2}&\Omega^{2} + y^{2} + 2x - \frac{2}{3}x^{2} + \Omega_{k} = 1.
\end{align}

The corresponding Jacobian matrix of the autonomous equation set in eq.~\eqref{eq:curved_odd_fed_set},
\begin{align}
\footnotesize
\mathcal{J} = 
\begin{pmatrix}- \frac{3 \Omega^{2} \left(\lambda + 2 \omega\right)}{2} - 4 x - 2 \Omega_{k} + 6 & 3 \Omega \left(\lambda + 3 \omega - x \left(\lambda + 2 \omega\right) + 1\right) & 3 - 2 x\\- \frac{\Omega \left(\lambda + 2\right)}{2} & - \frac{9 \Omega^{2} \left(\lambda + 2 \omega\right)}{2} - \frac{\lambda x}{2} - \frac{3 \omega}{2} - x - 2 z + \frac{3}{2} & - 2 \Omega\\- 4 \Omega_{k} & - 6 \Omega \Omega_{k} \left(\lambda + 2 \omega\right) & - 3 \Omega^{2} \left(\lambda + 2 \omega\right) - 4 x - 8 \Omega_{k} + 4\end{pmatrix}.
\end{align}

\begin{figure}[htp]
    \centering
    \begin{subfigure}[b]{0.46\textwidth}
        \includegraphics[width=\linewidth]{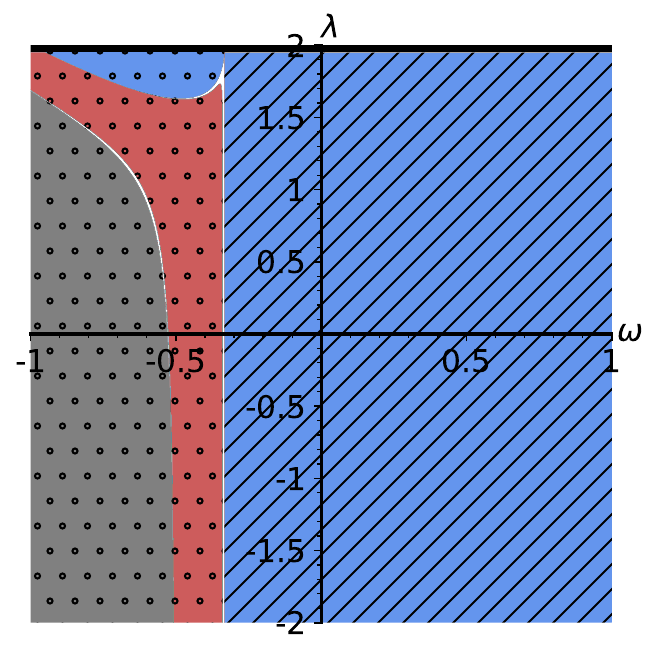}
    \caption{\label{fig:curved_C67}}
    \end{subfigure}
    \begin{subfigure}[b]{0.46\textwidth}
        \includegraphics[width=\linewidth]{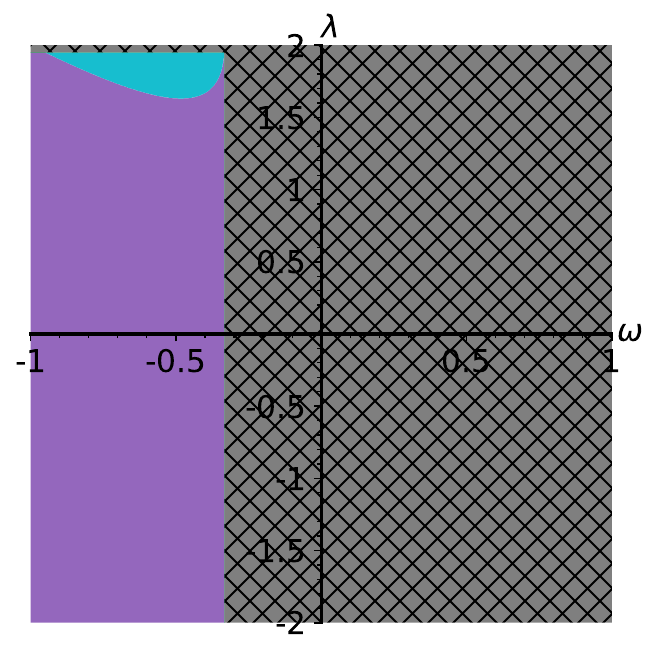}
    \caption{\label{fig:curved_C67_OmegaK}}
    \end{subfigure}
    \caption{\label{fig:curved_ol67}On the left-hand side, we give the phase space behavior analysis of critical points $C_{6}$ and $C_{7}$ of the Table \ref{tab:curved_x-Omega-hyper-criticalpoints-o2} with respect to equation of state parameters $\omega$ and $\lambda$ where we included the curvature contribution. White curves between regions represent bifurcation curves. Green: Source node, Blue: Saddle node, Red: Attractor node, Grey: Stable spiral (focus),  Black strip at the top of the figure shows $\lambda = -2$, the shaded regions with straight line ($\omega > -1/3$) represents the forbidden regions and the shaded regions with dots represents the accelerating regions. On the right-hand side, we give the same critical point's curvature characteristics whether the Universe is open ($k<0$) with purple color, closed ($k>0$) with cyan color or flat ($k=0$) with green color, the grey shaded area represents the forbidden region and unshaded area represents the accelerating region.}
\end{figure}

We again take the flux term ($y$) as zero without the loss of generality in our calculations but write it in our equation set for completeness. The explicit contribution of the flux term can be realized as mentioned previously by setting the equation of state variables $\omega$ and $\lambda$ to specific values matching to the one of flux. The autonomous equation set of the new variables therefore can be written as,
\begin{subequations}
\label{eq:curved_autonomous_original}
\begin{align}
&x' = \frac{\ddot{\phi}}{H^{2}} - x\frac{\dot{H}}{H^{2}}\\
&y' = -y\left(3 + \frac{\dot{H}}{H^{2}}\right)\\
&\Omega' = \Omega\left(\frac{1}{2}\frac{\dot{\rho}}{H\rho} + x - \frac{\dot{H}}{H^{2}}\right)\\
&\Omega'_{k} = -2\Omega_{k}\left(1 + \frac{\dot{H}}{H^{2}}\right).
\end{align}
\end{subequations}
Explicit dependence to the dynamical variables of the closed expressions in the above equation set is given as,
\begin{subequations}
    \begin{align}
        &\frac{\dot{H}}{H^{2}} = \Omega^{2} \left(3\omega + \frac{3\lambda}{2}\right) + 6y^2 - 3 + 2x + 2\Omega_{k}\\
        &\frac{\ddot{\phi}}{H^{2}} = \frac{3}{2}\left(\frac{\dot{H}}{H^{2}} + 1\right) + \frac{3\Omega^{2}}{2}\left(1 - \frac{\lambda}{2}\right)\\
        &\frac{\dot{\rho}}{H\rho} = -3(1 + \omega) - x\lambda
    \end{align}
\end{subequations}
In Table \ref{tab:curved_x-Omega-hyper-criticalpoints-o2} we give the linear stability analysis results with the inclusion of the curvature term $\Omega_{k}$. There are three critical points with non-zero curvature, $C_{1}$, $C_{6}$ and $C_{7}$. Critical point $C_{1}$ always shows a saddle node character with the curvature constant being $k<0$ coinciding with the open universe configuration. The equation of state parameter analysis for the points $C_{6}$ and $C_{7}$ is given with Figure \ref{fig:curved_C67}, for which it can be seen that in the allowed region the Universe is accelerating with the value of the curvature mostly favors $k < 0$ which can be seen in \ref{fig:curved_C67_OmegaK}, i.e. an open universe. The comparison of the two figures in \ref{fig:curved_ol67} shows that the attractor and stable spiral characteristics of the accelerating Universe only presents itself in the spatially open ($k < 0$) Universe and the critical point shows the saddle behavior in the spatially closed ($k > 0$) Universe.

\bgroup
\def\arraystretch{1.2}%
\begin{table}[htp]
\begin{tabular}{c | c | c | c | l | l}
\hline
\footnotesize{\textbf{\#}} & \footnotesize{\boldmath$x$} & \footnotesize{\boldmath{$\Omega$}} & \footnotesize{\boldmath{$\Omega_{k}$}} & \footnotesize{\textbf{Eigenvalues}} & \footnotesize{\textbf{Existence}}\\
\hline
\hline
$C_{1}$ & $0$ & $0$ & $1$ & $\tiny{\begin{array}{ll}&\hspace{-0.0cm}\lambda_{1} = -2\\&\hspace{-0.0cm}\lambda_{2} = 2\\&\hspace{-0.0cm}\lambda_{3} = \frac{-(3\omega + 1)}{2}\end{array}}$ & $\scriptsize{\begin{array}{ll}&\forall (\omega,\lambda)\end{array}}$ \\
\hline
$C_{2}$ & $\frac{3 + \sqrt{3}}{2}$ & $0$ & $0$ & $\tiny{\begin{array}{ll}&\hspace{-0.0cm}\lambda_{1} = -2-2\sqrt{3}\\&\hspace{-0.0cm}\lambda_{2} = -2\sqrt{3}\\&\hspace{-0.0cm}\lambda_{3} = \dfrac{-\lambda(3 + \sqrt{3}) - 2(3\omega + \sqrt{3})}{4}\end{array}}$ & $\scriptsize{\begin{array}{ll}&\forall (\omega,\lambda)\end{array}}$\\
\hline
$C_{3}$ & $\frac{3 - \sqrt{3}}{2}$ & $0$ & $0$ & $\tiny{\begin{array}{ll}&\hspace{-0.0cm}\lambda_{1} = -2+2\sqrt{3}\\&\hspace{-0.0cm}\lambda_{2} = 2\sqrt{3}\\&\hspace{-0.0cm}\lambda_{3} = \dfrac{-\lambda(3 - \sqrt{3}) - 2(3\omega - \sqrt{3})}{4}\end{array}}$ & $\scriptsize{\begin{array}{ll}&\forall (\omega,\lambda)\end{array}}$\\
\hline
$C_{4,5}$ & $\frac{g(\omega,\lambda)}{\lambda + 2\omega}$ & $\pm\sqrt{\frac{h(\omega,\lambda)}{(\lambda + 2\omega)^{2}}}$ & $0$ & $\tiny{\begin{array}{ll}&\hspace{-0.0cm}\lambda_{1} = \frac{g(\omega,\lambda)^{2} - (3\omega+1)(\omega-1)}{\lambda + 2\omega}\\&\hspace{-0.0cm}\lambda_{2} = \dfrac{g(\omega,\lambda)^{2} - 3(\omega - 1)^{2}}{2(\lambda + 2\omega)}\\&\hspace{-0.0cm}\lambda_{3} = \dfrac{g(\omega,\lambda)^{2} - 3(\omega - 1)^{2} + 2(\lambda + 2)}{\lambda + 2\omega}\end{array}}$ & $\scriptsize{\begin{array}{ll}&\lambda \ne -2\omega\\&h(\omega,\lambda) \ge 0\end{array}}$\\
\hline
$C_{6,7}$ & $-\frac{3\omega + 1}{\lambda - 2}$ & $\pm\frac{2\sqrt{-(\omega + \frac{1}{3})}}{\lambda - 2}$ & $\frac{g(\omega,\lambda)^{2} - (3\omega + 1)(\omega - 1)}{(\lambda - 2)^{2}}$ & $\tiny{\begin{array}{ll}&\hspace{-0.0cm}\lambda_{1} = 2\\&\hspace{-0.0cm}\lambda_{2} = \dfrac{g(\omega,\lambda) - f(\omega,\lambda)}{2 - \lambda}\\&\hspace{-0.0cm}\lambda_{3} = \dfrac{g(\omega,\lambda) + f(\omega,\lambda)}{2 - \lambda}\end{array}}$ & $\scriptsize{\begin{array}{ll}&\lambda \ne 2\\&\omega \le -1/3\end{array}}$\\
\hline
$C_{8,9}$ & $-\frac{3(\omega + 1)}{\lambda - 2}$ & $\pm\sqrt{\frac{2}{\lambda - 2}}$ & $0$ & $\tiny{\begin{array}{ll}&\hspace{-0.0cm}\lambda_{1} = -2\\&\hspace{-0.0cm}\lambda_{2} = \dfrac{-3(\lambda + 2\omega) - \sqrt{3\Xi}}{2(\lambda - 2)}\\&\hspace{-0.0cm}\lambda_{3} = \dfrac{-3(\lambda + 2\omega) + \sqrt{3\Xi}}{2(\lambda - 2)} \end{array}}$ & $\scriptsize{\begin{array}{ll}&\lambda \ne 2\\&\Xi \ge 0\end{array}}$\\
\hline
\hline
\multicolumn{5}{c}{$\small{\begin{array}{ll}&\Xi(\omega,\lambda) \equiv -4\lambda^{3} - 24(\lambda^{2}\omega + \lambda\omega^{2}) + 11\lambda^{2} + 60\omega(\lambda + \omega) - 8\lambda + 16\end{array}}$}\\
\hline
\hline
\end{tabular}
\caption{\label{tab:curved_x-Omega-hyper-criticalpoints-o2}Critical points of autonomous system in eq.~\eqref{eq:curved_autonomous_original}}
\end{table}
\egroup

\section{Results and Discussion}
One of the most elegant approaches to studying the cosmological equations that describe the dynamics of a homogeneous and isotropic Universe is to transform them into dynamical systems. The equations of such a maximally symmetric cosmological setting form systems of ordinary differential equations that can be brought into the form of autonomous systems. This enables a quantitative comprehension of the cosmic dynamics provided by the models under study to be obtained through the application of strong analytical and numerical approaches. In this manuscript, we incorporate these methods in the context of $O(D, D)$-complete Friedmann cosmology. In doing this, one of our primary aims is to study the dynamical system approach for comparing the $O(D, D)$-extended Friedmann cosmology with the Chameleon cosmology to see if the $O(D, D)$-symmetric system has possible phenomenological applications to both early and late universe problems. It is expected at the onset of analysis that the $O(D, D)$-extended cosmological system has richer possibilities, and this is realized in the form of a hyperboloid of two sheets in the phase space construction. On the other hand, the two-sheeted hyperboloid can be put into the form of a sphere, and we prefer to consider such a compact region in phase space for our analysis. In the cosmologically interesting cases of scalar fields, radiation, and non-relativistic matter there is at least an attractor point and a source point. We also considered the physical meaning and importance of these critical points for whether they lead to acceleration (or deceleration) of the universe. We compared with the similar analysis of the Chameleon cosmology and saw that accelerating regions in both models coincide with the flux and scalar field potential domination in $O(D, D)$ and Chameleon models, respectively. Furthermore, in $O(D, D)$ setting, when the flux contribution comes to domination and eventually reaches the attractor node, the effective equation of state parameter relaxes to $-1/3$; but in the Chameleon setting, when the scalar field potential starts to dominate, the effective equation of state parameter goes to $-1$. 

The reason for the comparison with the chameleon cosmic model was that we expected that the $O(D, D)$-extended system would give rise to similar, or even richer, dynamics when analyzed and compared in terms of dynamical systems, and the aforementioned facts confirm this expectation. This would allow $O(D, D)$-symmetric systems to provide a physically sound alternative to general relativity in addressing cosmic problems, instead of modifications suggested on pure phenomenological ground.

We also introduced a spatial curvature term ($\Omega_{k}$) into our calculations and investigated the acceleration characteristics associated with a non-zero curvature contribution. The findings revealed that the Universe predominantly experiences an accelerating phase in the open Universe ($k < 0$), with the critical point acting as either an attractor or a stable spiral (focus). In a limited range of $\omega$ and $\lambda$ values, the critical point exhibited saddle behavior within the accelerating region for closed Universe values ($k > 0$). A graphical representation of these results is provided in Figure \ref{fig:curved_ol67}.

There are some other important cosmological problems in this setting, such as checking the generation of scale-free cosmological perturbations, properties of a possible stochastic gravitational wave background, and nucleosynthesis effects. Furthermore, the coupling of the dilaton to matter part in a certain symmetry-dictated way can lead to observable effects on the structure of compact objects like neutron stars. These are possible further research problems that we are working on.

\section*{Acknowledgement}
Authors are supported by The Scientific and Technological Research Council of T\"{u}rkiye (T\"{U}B\.{I}TAK) through the ARDEB 1001 project with grant number 121F123. We would also like to thank the anonymous referee for contributing to clarifying some points with relevant and important suggestions.
\bibliographystyle{apsrev4-2}
\bibliography{biblio}

\end{document}